\documentclass[superscriptaddress,showpacs,amssymb,10pt,reprint,aps,prd,longbibliography,nofootinbib,floatfix]{revtex4-2}

\usepackage{graphicx,epsfig,amssymb,times} 
\usepackage{amsmath,amsfonts}
\usepackage{bm}
\usepackage{epstopdf}
\usepackage{hyperref}
\usepackage[usenames]{color}   
\usepackage[dvipsnames]{xcolor}
\usepackage[titletoc]{appendix}
\usepackage{subfigure}
\usepackage{graphicx}

\usepackage[normalem]{ulem}
 
\definecolor{coolblack}{rgb}{0.0, 0.18, 0.39}
\definecolor{darkred}{rgb}{0.5,0,0}
\definecolor{darkgreen}{rgb}{0,0.5,0}
\definecolor{darkblue}{rgb}{0,0,0.5}
\definecolor{lapislazuli}{rgb}{0.15, 0.38, 0.61}
\definecolor{venetianred}{rgb}{0.78, 0.03, 0.08}
\definecolor{bleudefrance}{rgb}{0.19, 0.55, 0.91}
\definecolor{dogwoodrose}{rgb}{0.84, 0.09, 0.41}
\hypersetup{colorlinks=true, citecolor=darkblue, linkcolor=darkblue, 
urlcolor = darkblue}

\begin{document}

\title{\large Sufficient conditions for unbounded superradiance in black hole spacetimes sourced by nonlinear electrodynamics}

	\author{Marco A. A. de Paula}
	\email{marco.paula@icen.ufpa.br}
	\affiliation{Programa de P\'os-Gradua\c{c}\~{a}o em F\'{\i}sica, Universidade Federal do Par\'a, 66075-110, Bel\'em, Par\'a, Brazil.}
		
	\author{Luiz C. S. Leite}
	\email{luiz.leite@ifpa.edu.br}
	\affiliation{Campus Altamira, Instituto Federal do Par\'a, 68377-630, Altamira, Par\'a, Brazil.}
	
	\author{Lu\'is C. B. Crispino}
	\email{crispino@ufpa.br}
	\affiliation{Programa de P\'os-Gradua\c{c}\~{a}o em F\'{\i}sica, Universidade Federal do Par\'a, 66075-110, Bel\'em, Par\'a, Brazil.}

\begin{abstract}

Recently, it has been reported that black holes (BHs) sourced by nonlinear electrodynamics (NED) can trigger unbounded superradiance, i. e., the total absorption cross section --- the ratio between the absorbed flux and the incident flux of the wave --- can be negative and unbounded from below. Considering the propagation of massive charged test scalar fields in the vicinity of electrically charged BH solutions based on the NED framework, we derive sufficient conditions to have an unbounded superradiant regime. We also discuss some possible general conditions that the NED electrical source associated with the geometry has to satisfy to trigger an unbounded superradiance. Our results apply to a broad family of electrically charged BH solutions (regular and singular) derived in the framework of NED minimally coupled to general relativity.

\end{abstract}

\date{\today}
	
\maketitle

\section{Introduction}

Since the pioneering works of Max Born and Leopold Infeld in the mid-1930s~\cite{B1933,BI1934,B1934}, efforts have been made to fathom the applications and implications of nonlinear electrodynamics (NED) models. For instance, there is NED research carried out in the quantum regime~\cite{HE1936,RPM2016,A2021,EK1935,ATLAS2017,ATLAS2019,PVLAS2020,D2003}, string/M-theory~\cite{FT1985,SW1999,A2000}, and cosmology~\cite{GB2000,VAL2002,NBS2004,LC2008,K2015,K2016,MT2022}. Furthermore, in 1998, Eloy Ayón-Beato and Alberto García (ABG for short) realized that it is possible to derive black hole (BH) solutions without a curvature singularity, i.e., regular BH (RBH) geometries, by using NED models~\cite{ABG1998}. Since then, several charged RBH spacetimes based on the NED framework have been proposed (see, e.g., Refs.~\cite{DPS2021,KAB2023} and references therein).

The standard method for exploring the imprints of NED models in the context of BH geometries is tracking down the main effects of the nonlinear electromagnetic field. We can achieve it by using different strategies. One can, e.g., investigate the stability of such geometries under electromagnetic and gravitational perturbations~\cite{MS2003,NYS2020,DG2022,NY2022} by considering the appropriate equations satisfied by the Faraday tensor [see, e.g., Eq.~\eqref{EFTC1}]. Another possibility is quantifying the effects associated with the geometry followed by photons~\cite{P1970,GDP1981,MN2000} on some optical phenomena, such as shadows and gravitational lensing~\cite{EE2006,SS20192,SS2019,SSO2019,AA2020,MP2023,SV2023}. Alternatively, we can study the superradiance phenomenon in NED-based RBH geometries~\cite{MAP2024c,MAP2024b,WL2024}. Moreover, there are also works investigating the causality properties of NED-sourced spacetimes~\cite{SPL2016,TS2023,RW2024,MS2024,MAP2024d}.

For charged scalar fields in the background of static and spherically symmetric spacetimes, there is a coupling between the field charge $q$ and the radial electrostatic potential $\phi(r)$. In this context, for a scalar field with frequency $\omega$ satisfying
\begin{equation}
\label{eq1}\omega < q\phi_{+},
\end{equation}
where $\phi_{+} \equiv \phi(r_{+})$ and $r_{+}$ is the event horizon radius, we can have a superradiant scattering, i.e., the (charged) scalar wave can be scattered by the central object with more energy than it originally had, by carrying away energy from the BH~\cite{B1973,G1975,NS1976,BC2016,BR2016,BCP2021}. To trigger the superradiant instability we need two ingredients: (i) the existence of a trapping potential well outside the BH and (ii) the existence of superradiant amplification of the charged field. The trapping potential can be provided by different mechanisms, e.g., presenting a boundary to the spacetime, as in anti--de Sitter geometries~\cite{CD2004}, or massive bosonic fields in the vicinity of spinning BHs~\cite{ZE1979,SD1980,SD2013}.

In the absence of trapping mechanisms, such as a cavity~\cite{DHH2013,DPW2015,DM2018}, the Reissner-Nordstr\"om (RN) BH geometry cannot present a superradiant instability under charged massive scalar wave perturbations~\cite{SH2012}. Yet, for electrically charged NED-sourced spacetimes, the picture is different. Recently, by investigating the behavior of a charged massive scalar field in the background of an ABG RBH, it has been reported that the ABG RBH can present superradiant instability in a certain parameter range~\cite{MAP2024b}. Other RBHs based on the NED framework can show similar behavior~\cite{ZXZ2024}. It was also reported in Ref.~\cite{SH2024} that it is possible to find a sufficient condition to trigger superradiant instability in the background of charged BHs. The NED theory also introduces nontrivial features into the behavior of the absorption cross section (ACS).

The ACS is the ratio between the absorbed flux and the incident flux of the wave~\cite{FHM1988} and has been studied in several scenarios over the last 50 years (see, e.g., Refs.~\cite{U1976,OCH2011,CB2014,LBC2017,LDC2017,LDC2018,BC2019,XBC2021,MC2014,SBP2017,S2017,PLC2020} and references therein). The behavior of the ACS in the low-frequency regime can be mapped into a $(\mu M/qQ, Q/M)$ plot called by absorption parameter space. In Ref.~\cite{MAP2024c}, a specific NED model has been considered, namely the ABG RBH derived in Ref.~\cite{ABG1998}. In Ref.~\cite{MAP2024c}, it has been shown that, for the ABG RBH, there are three possibilities: unbounded absorption, unbounded superradiance, and bounded superradiance. For unbounded absorption, the total ACS diverges from above without presenting superradiance. Moreover, for impinging waves within the superradiant regime, the total ACS diverges from below, for unbounded superradiance, and becomes negative, as a consequence of the superradiant scattering, but finite when bounded superradiance occurs. This is notably different from the RN case, where instead of an unbounded superradiance region, we have a scenario of bounded absorption.

It is important to emphasize that superradiant instability and unbounded superradiance are distinct phenomena. Unbounded superradiance is associated with the divergence of the ACS from below during the superradiant scattering of a wave impinging upon the BH from infinity. In contrast, superradiant instability is related to the formation of superradiant bound states, which correspond to bound states formed in the superradiant regime. Therefore, these two concepts cannot be trivially exchanged.

Here, we aim to provide sufficient conditions for the existence of an unbounded superradiant regime in the background of NED-based RBH geometries, considering charged massive scalar waves. To verify the accurateness of the sufficient conditions, we consider some well-known NED-based electrically charged RBH solutions presented in the literature [namely, the solutions of ABG~\cite{ABG1998}, Irina Dymnikova (ID)~\cite{D2004}, and Balart and Vagenas (BV)~\cite{BV2014}]. We show the structural function, metric function, and radial electrostatic potential of these spacetimes in Appendix~\ref{apx}.

Moreover, we also discuss general conditions that the electrical NED source itself associated with the BH geometries has to satisfy to guarantee an unbounded superradiant regime. The remainder of this work is organized as follows. In Sec.~\ref{sec:frame}, we detail the framework used to obtain the BH geometries. The dynamics of charged massive scalar fields is considered in Sec.~\ref{sec:csf}, and our main results are addressed in Sec.~\ref{sec:rd}. Throughout this paper, we use the natural units, for which $G = c = \hbar = 1$, and the metric signature ($-,+,+,+$).

\section{Framework}\label{sec:frame}

The action that describes GR minimally coupled to NED, in the so-called $F$ framework, can be written as 
\begin{equation}
\label{S}\mathrm{S} = \dfrac{1}{16\pi}\int d^{4}x \left[R-\mathcal{L}(F) \right]\sqrt{-g},
\end{equation}
in which $R$ is the Ricci scalar, $\mathcal{L}(F)$ is a gauge-invariant electromagnetic Lagrangian density, and $g$ is the determinant of the metric tensor $g_{\mu\nu}$. The Maxwell scalar $F$ is given by
\begin{equation}
\label{maxwellscalar}F \equiv F_{\mu\nu}F^{\mu\nu},
\end{equation}
with $F_{\mu\nu} = 2\nabla_{[\mu}A_{\nu]}$ being the standard electromagnetic field strength, which satisfies
\begin{equation}
\label{EFTC1}\nabla_{\mu}\left(\mathcal{L}_{F}F^{\mu\nu}\right)  = 0 \ \ \text{and} \ \ \nabla_{\mu}\star F^{\mu\nu} = 0,
\end{equation}  
where $\mathcal{L}_{F} = \partial \mathcal{L}/\partial F$, $\star F^{\mu\nu}$ is the dual electromagnetic field tensor, and the quantity $A_{\mu}$ is the four-vector potential. The corresponding field equations are given by
\begin{equation}
\label{E-NED_F}G_{\mu}^{\ \nu} = 8\pi T_{\mu}^{\ \nu} = 2\left[\mathcal{L}_{F}F_{\mu\alpha}F^{\nu\alpha}-\dfrac{1}{4}\delta_{\mu}^{\ \nu}\mathcal{L}(F)\right].
\end{equation}

The field system given by Eqs.~\eqref{S}-\eqref{E-NED_F} reduces to the standard Maxwell theory in the weak field limit if
\begin{equation}
\label{maxasy}\mathcal{L}(F) \rightarrow F \ \  \text{and} \  \ \mathcal{L}_{F} \rightarrow 1,
\end{equation}
for small $F$. The conditions~\eqref{maxasy} are the Maxwell asymptotic conditions. Moreover, assuming the weak field regime, it is not possible to obtain static and spherically symmetric solutions with a regular center and a nonzero electric charge from a single Lagrangian density $\mathcal{L}(F)$~\cite{B2001}. 

In order to obtain exact electrically charged RBH solutions in NED, we typically introduce a secondary framework. This can be done by introducing an auxiliary antisymmetric tensor $\textbf{P}$, with covariant components $P_{\mu\nu}$. In this context, we rewrite Eqs.~\eqref{S}-\eqref{E-NED_F} in terms of $P_{\mu\nu}$ by defining a structural function $\mathcal{H}(P)$ via Legendre transformation~\cite{HGP1987}, namely
\begin{equation}
\label{LT_H}\mathcal{H}(P) \equiv 2F\mathcal{L}_{F} - \mathcal{L}(F),
\end{equation}
with the scalar $P$ and the tensor $P_{\mu\nu}$ being defined as
\begin{equation}
\label{AT}P \equiv P_{\mu\nu}P^{\mu\nu} \ \ \text{and} \ \ P_{\mu\nu} \equiv \mathcal{L}_{F}F_{\mu\nu},
\end{equation}
respectively. By using Eqs.~\eqref{LT_H} and~\eqref{AT}, we can show that
\begin{equation}
\label{FPDUALITY}P = (\mathcal{L}_{F})^{2}F\ \ \text{and} \ \ \mathcal{H}_{P}\mathcal{L}_{F} = 1,
\end{equation}
where $\mathcal{H}_{P} \equiv \partial\mathcal{H}/\partial P$. In the $P$ framework, we obtain the Maxwell electrodynamics in the far field if $\mathcal{H}(P) \rightarrow P$ and $\mathcal{H}_{P} \rightarrow 1$, for small $P$. Although RBH solutions obtained in the $P$ framework can be formally derived in the $F$ framework using a suitable nonuniform variational method~\cite{DGT2015}, it is easier to use the $P$ framework than the $F$ framework for such solutions, at least from the algebraic point of view.

This is related to the following issue. Any well-behaved electrically charged RBH solution obtained in the $P$ framework with Maxwell asymptotics, requires (at least two) different Lagrangian densities $\mathcal{L}(F)$ in distinct parts of the spacetime~\cite{B2001}. Even electrically charged BH solutions obtained in the $F$ framework that are functions of only a single Lagrangian density $\mathcal{L}(F)$ are trickier to work with, due to the nontrivial behavior of the conservation equation for $F_{\mu\nu}$, given by $\nabla_{\mu}\left(\mathcal{L}_{F}F^{\mu\nu}\right)  = 0$. In the $P$ framework, the conservation equation for the tensor $P_{\mu\nu}$ is merely $\nabla_{\mu}P^{\mu\nu} = 0$, which is notably simpler.

Therefore, since we are mainly interested in electrically charged RBH spacetimes with a regular core and a Maxwell asymptotic behavior at infinity, we work here with the $P$ framework, where these spacetimes can be derived straightforwardly. For completeness, in the Appendix~\ref{Appendix} we extend our main results obtained in the $P$ framework to the $F$ framework and discuss them. For more details about these frameworks see, e.g., Refs.~\cite{HGP1987,B2001,D2004,BV2014,DGT2015}, and references therein. 

One can solve the field equations providing an ansatz for the line element. We choose a static and spherically symmetric spacetime, for which the line element is given by
\begin{equation}
\label{LE}ds^{2} = -f(r)dt^{2}+f(r)^{-1}dr^{2}+r^{2}d\Omega^{2},
\end{equation}
where $d\Omega^{2} = d\theta^{2}+\sin^{2}\theta d\varphi^{2}$ is the line element of a unit two-sphere and $f(r)$ is the metric function, given by
\begin{equation}
\label{MF}f(r) = 1 - \dfrac{2\mathcal{M}(r)}{r}.
\end{equation}
The function $\mathcal{M}(r)$ is determined by the field equations, and from its asymptotic behavior we can obtain the total mass $\mathcal{M}(r \rightarrow \infty) = M$ of the central object~\cite{FW2016}. The ansatz given by Eq.~\eqref{LE} is enough for our purposes if we require a correspondence with general relativity~\cite{SR2018}, which is the case.

\begin{figure}[!htbp]
\begin{centering}
    \includegraphics[width=1\columnwidth]{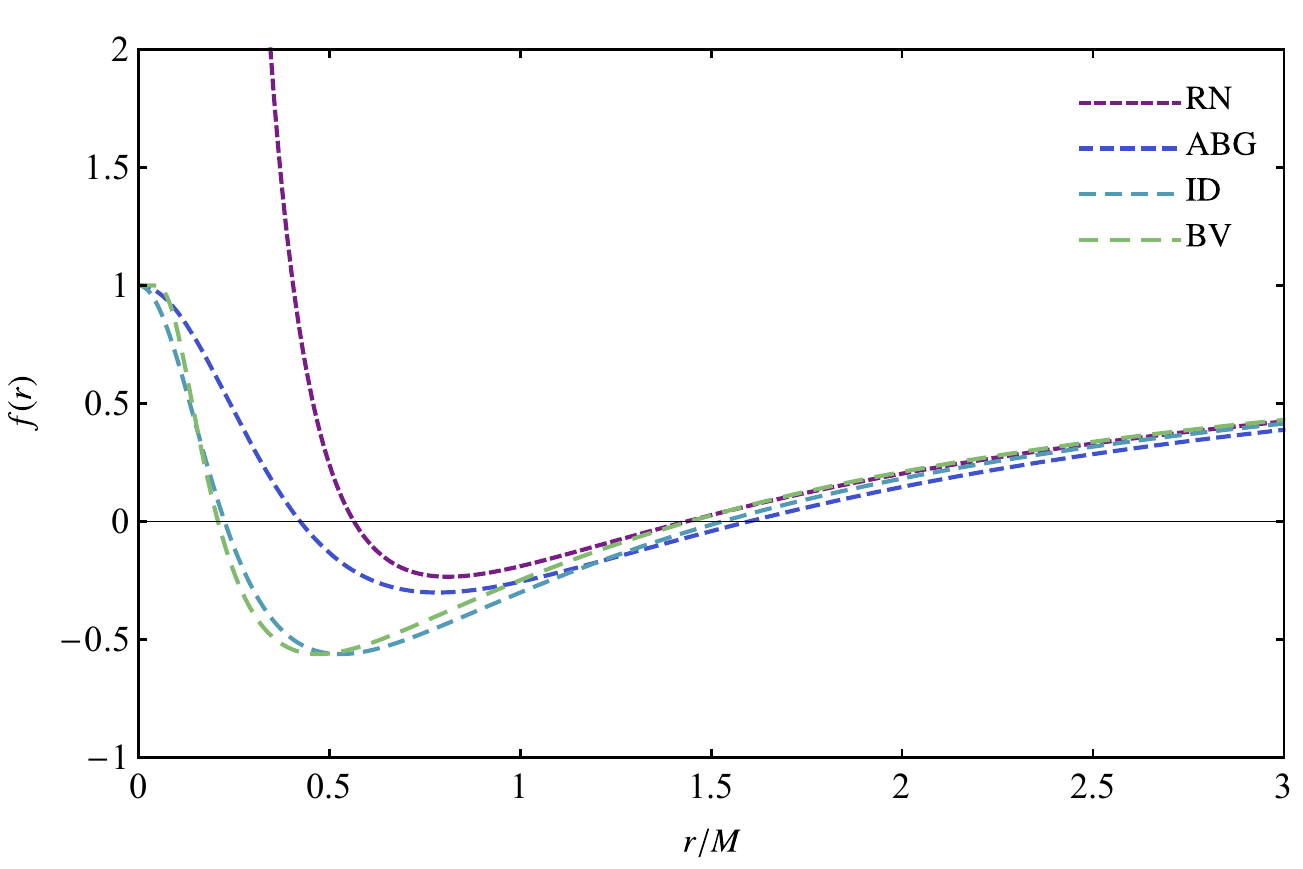}
    \caption{Metric function $f(r)$ of RN, ABG, ID, and BV geometries, considering $\alpha = 0.8$, as functions of $r/M$.}
    \label{metricfunctions}
\end{centering}
\end{figure}

In Fig.~\ref{metricfunctions}, we display the metric function of the RN, ABG, ID, and BV BH solutions. We use the ratio $\alpha \equiv Q/Q_{\rm{ext}}$, where $Q_{\rm{ext}}$ is the extremal value of the charge, and we set $\alpha = 0.8$ for all the solutions shown in the figure. As we can see, the geometries share a similar causal structure. For $\alpha < 1$, we have an inner horizon, $r_{-}$, and an event horizon, $r_{+}$. On the other hand, in the extremely charged scenario, $r_{-} = r_{+} \equiv r_{\rm{ext}}$, where $r_{\rm{ext}}$ is the extremal horizon. Moreover, the uncharged limit (Schwarzschild spacetime) is properly recovered when $\alpha = 0$. In this paper, we are only interested in BH solutions, for which $0 \leq \alpha \leq 1$. Notice that $r_{\rm{ext}}$ and $Q_{\rm{ext}}$ can be obtained by solving $f(r) = 0$ and $df(r)/dr = 0$, simultaneously, leading to $ Q^{\rm{RN}}_{\rm{ext}} = M$, $Q^{\rm{ABG}}_{\rm{ext}} = 0.6341M$, $Q^{\rm{ID}}_{\rm{ext}} = 1.073M$, and $Q^{\rm{BV}}_{\rm{ext}} = 1.213M$.

For purely electrically charged structures in the $P$ framework, the only non-null components of the tensor $P_{\mu\nu}$ are given by $P_{01}$ and $P_{10}$, which are related by
\begin{equation}
P_{01} = -P_{10} = -\dfrac{Q}{r^{2}},
\end{equation}
where $Q$ is the electric charge. Accordingly,
\begin{equation}
\label{ms}P = -\dfrac{2Q^{2}}{r^{4}}.
\end{equation}
Moreover, we know that the electric field is given by $F_{01} = -E(r)$. Therefore, by using Eqs.~\eqref{AT} and~\eqref{FPDUALITY}, it follows that
\begin{equation}
\label{electric}E(r) = \dfrac{Q\mathcal{H}_{P}}{r^{2}}.
\end{equation}

The covariant components of the electromagnetic four-potential are given by 
\begin{equation}
	\label{VP_ABG}A_{\mu}=(-\phi(r),0,0,0),
\end{equation}
in which $\phi(r)$ is the radial electrostatic potential. We can obtain a generic expression for the radial electrostatic potential by using $\phi(r) = -\int_{\mathcal{O}}^{r}\textbf{E}\cdot d\textbf{l}$ and Eq.~\eqref{electric}, and hence
\begin{equation}
\label{EP}\phi(r) = -\int_{\infty}^{r}\dfrac{Q\mathcal{H}_{P}}{x^{2}}dx,
\end{equation}
where, for convenience, we placed the reference point $\mathcal{O}$ of the radial electrostatic potential at infinity.

\begin{figure}[!htbp]
\begin{centering}
    \includegraphics[width=1\columnwidth]{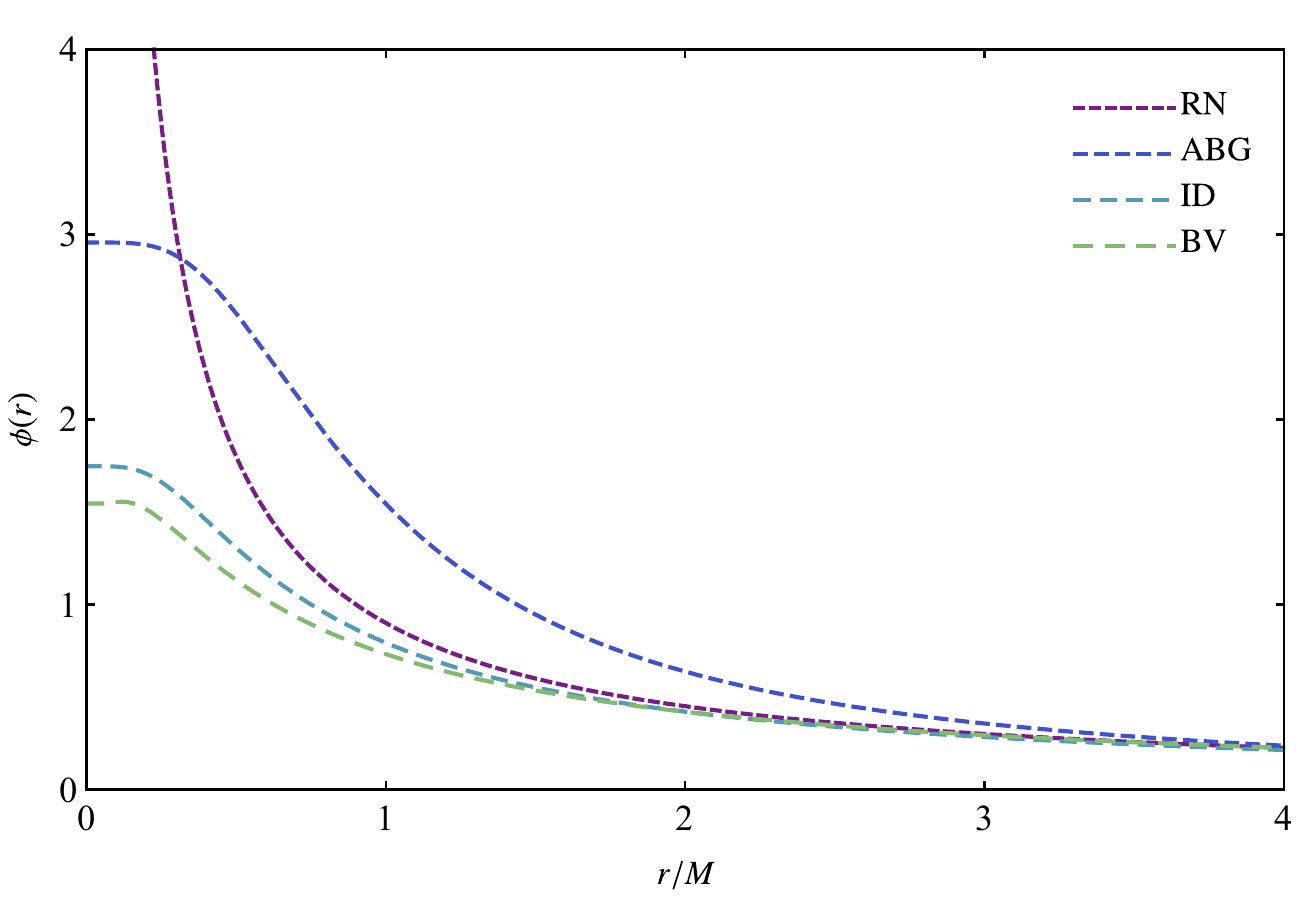}
    \caption{Electrostatic potential $\phi(r)$ of RN, ABG, ID and BV geometries, considering $\alpha = 0.8$, as functions of $r/M$.}
    \label{potentials}
\end{centering}
\end{figure}

In Fig.~\ref{potentials}, we present the electrostatic potential of the RN, ABG, ID, and BV BH solutions. As we can see, the electrostatic potential  of regular geometries is finite at the coordinate center ($r = 0$), in contrast to the RN spacetime. Moreover, the electrostatic potential of regular spacetimes tends to the RN electrostatic potential in the weak field regime.

\section{Massive charged scalar waves}\label{sec:csf}

The dynamics of a charged massive test scalar field $\Phi$ in the background of curved spacetimes is governed by
\begin{equation}
\label{KG}\left(\nabla_{\nu}-iqA_{\nu}\right)\left(\nabla^{\nu}-iqA^{\nu}\right)\Phi - \mu^{2}\Phi = 0,
\end{equation}
where $q$ and $\mu$ are the charge and mass of the scalar field, respectively. As a solution to Eq.~\eqref{KG}, we can write $\Phi$ as
\begin{equation}
\label{PHI}\Phi\equiv \sum_{l}^{\infty} \frac{\Psi_{l}(r)}{r}P_{l}(\cos\theta)e^{-i\omega t},
\end{equation}
in which $\Psi_{l}(r)$ are radial functions and $P_{l}(\cos\theta)$ are the Legendre polynomials, with $l$ and $\omega$ being the angular momentum and frequency of the scalar wave, respectively. By inserting Eq.~\eqref{PHI} into Eq.~\eqref{KG}, we can obtain a radial equation, namely
\begin{equation}
\label{RE} \frac{d^{2}}{dr_{\star}^{2}}\Psi_{l}-V(r)\Psi_{l}=0,
\end{equation}
where $r_{\star}$ is the tortoise coordinate, defined by $dr_{\star} \equiv dr/f(r)$, and the function $V(r)$ is defined as
\begin{equation}
\label{EffP} V(r) \equiv f(r)\left[\mu^{2}+\frac{f^{\prime}(r)}{r}+ \dfrac{l(l+1)}{r^{2}}\right]-\tilde{\omega}^{2},
\end{equation}
where the prime symbol $\prime$ denotes differentiation with respect to the radial coordinate $r$ and $\tilde{\omega} \equiv \omega - q\phi(r)$.

In the typical absorption/scattering problem, the scalar field perturbation satisfies the following boundary conditions~\cite{AS}
\begin{equation}
\label{BC}\Psi_{\omega l}\sim\begin{cases}
T_{\omega l}e^{-ir_{\star}\zeta}, & r_{\star}\rightarrow -\infty\\
I_{\omega l}e^{-ir_{\star}\kappa}+R_{\omega l}e^{ir_{\star}\kappa}, & r_{\star}\rightarrow+\infty
\end{cases},
\end{equation}
where $\zeta\equiv\omega-q\phi_{+}$, $\kappa\equiv\sqrt{\omega^{2}-\mu^{2}}$, and $r_{+}$ is the event horizon radius, given by the largest positive root of $f(r) = 0$. The quantities $T_{\omega l}$, $I_{\omega l}$, and $R_{\omega l}$ are complex coefficients. Recall also that for the absorption/scattering problem, we require that $\omega^{2} > \mu^{2}$, since we are searching for unbounded modes. These boundary conditions are consistent with a wave impinging on the BH from infinity that, when interacting with the effective potential, is partially transmitted to the BH and partially reflected to infinity. That is different from the boundary conditions for studying (quasi)bound states, where we have a purely ingoing wave (the perturbation is localized near the BH).

The transmission and reflection coefficients are defined as
\begin{equation}
\label{transreflec}|\mathcal{T}_{\omega l}|^{2} \equiv \dfrac{|T_{\omega l}|^{2}}{|I_{\omega l}|^{2}} \ \ \ \text{and} \ \ \  |\mathcal{R}_{\omega l}|^{2} \equiv \dfrac{|R_{\omega l}|^{2}}{|I_{\omega l}|^{2}},
\end{equation}
respectively. From the conservation of the flux, we find
\begin{equation}
\label{CF}|\mathcal{R}_{\omega l}|^{2}+\frac{\zeta}{\kappa}|\mathcal{T}_{\omega l}|^{2}=1.
\end{equation}
Notice that when $q\phi_{+}>0$ and $\omega < q\phi_{+}$, we obtain that $\zeta<0$. In this case, we have an outgoing wave at $r_{+}$, which is amplified due to superradiance. The threshold for superradiant scattering is then given by
\begin{equation}
\label{criticalfreq}\omega_{c} = q\phi_{+}.
\end{equation}

For a plane wave incident upon a spherically symmetric BH, the ACS can be expanded in partial waves as
\begin{equation}
\label{TACS}\sigma=\sum_{l=0}^{\infty}\sigma_{l},
\end{equation}
where the partial ACS is
\begin{align}
\label{PACS}
\sigma_{l} &= \frac{\pi}{\kappa^3}(2l+1) (\omega - \omega_c) |\mathcal{T}_{\omega l}|^{2} . 
\end{align}
For superradiant modes (i.e., $0 < \omega < \omega_c = q \phi_+$), $\sigma_l$ takes negative values. In the limit $\omega \rightarrow \mu$ (from above), the momentum of the wave tends to zero, $\kappa \rightarrow 0$, and $\sigma_l$ will diverge in this limit, unless $\lim_{\omega \rightarrow \mu} |\mathcal{T}_{\omega l}|^{2} / \kappa^3$ exists. The divergent case is called \emph{unbounded} and the finite case \emph{bounded}~\cite{MAP2024c}.

\section{Results and discussions}\label{sec:rd}

In Ref.~\cite{MAP2024c}, considering charged massive test scalar fields, it was reported that it is possible to have an unbounded superradiance regime in the background of the ABG RBH for a given parameter range~\cite{ABG1998}. Essentially, we can find such situations in the ABG RBH because the strength of the vector potential at the event horizon is higher than that in the RN case. Consequently, under certain circumstances, the total absorption cross section can be negative and unbounded from below, in contrast with the RN case. In this section, we present sufficient conditions for having an unbounded superradiance regime in electrically charged NED-based BH spacetimes, covering a wide range of NED models.

\subsection{Effective potential}

We first discuss the role of the effective potential by recalling the notion of ``critical BH charges''~\cite{MAP2024c}. Considering
\begin{align}
\label{newpotef} U(r) \equiv & -V(r)|_{\mu = \omega},
\end{align}
the critical BH charge $Q_{c}$ are the values of $Q$ satisfying
\begin{equation}
\label{critcharge}U(r_{\rm{max}},Q_{c}) = 0,
\end{equation}
where $r_{\rm{max}}$ is the radial coordinate at which the effective potential achieves its local maximum value. In the situations where $Q < Q_{c}$, we have an unbounded regime for the absorption cross section, i.e., the cross section diverges in the limit $\mu M \rightarrow \omega M$. In turn, for $Q > Q_{c}$, we have a bounded regime. Depending on the parameter space range, the unbounded and bounded regimes can exhibit superradiance. 

\begin{figure}[!htbp]
\begin{centering}
    \includegraphics[width=1\columnwidth]{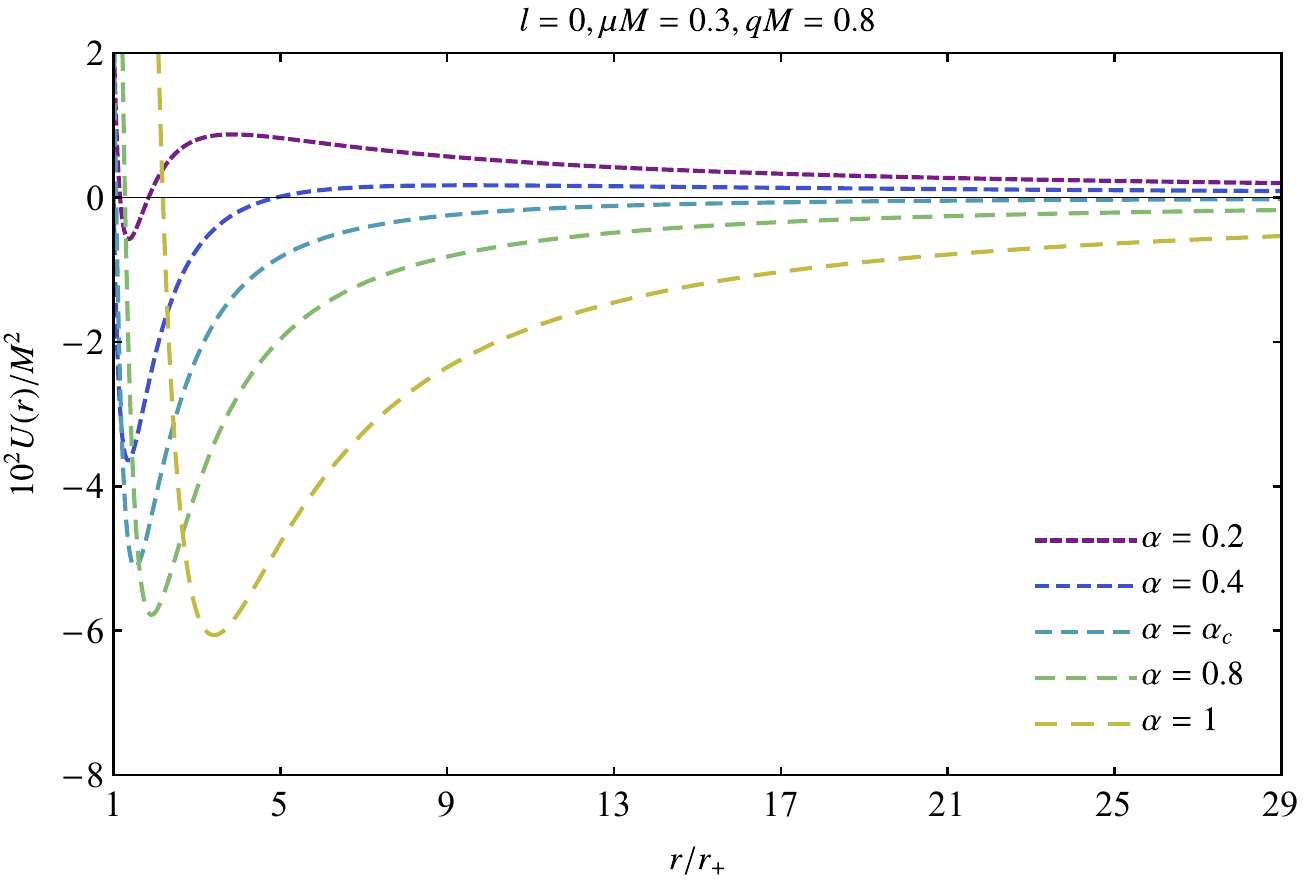}
    \includegraphics[width=1\columnwidth]{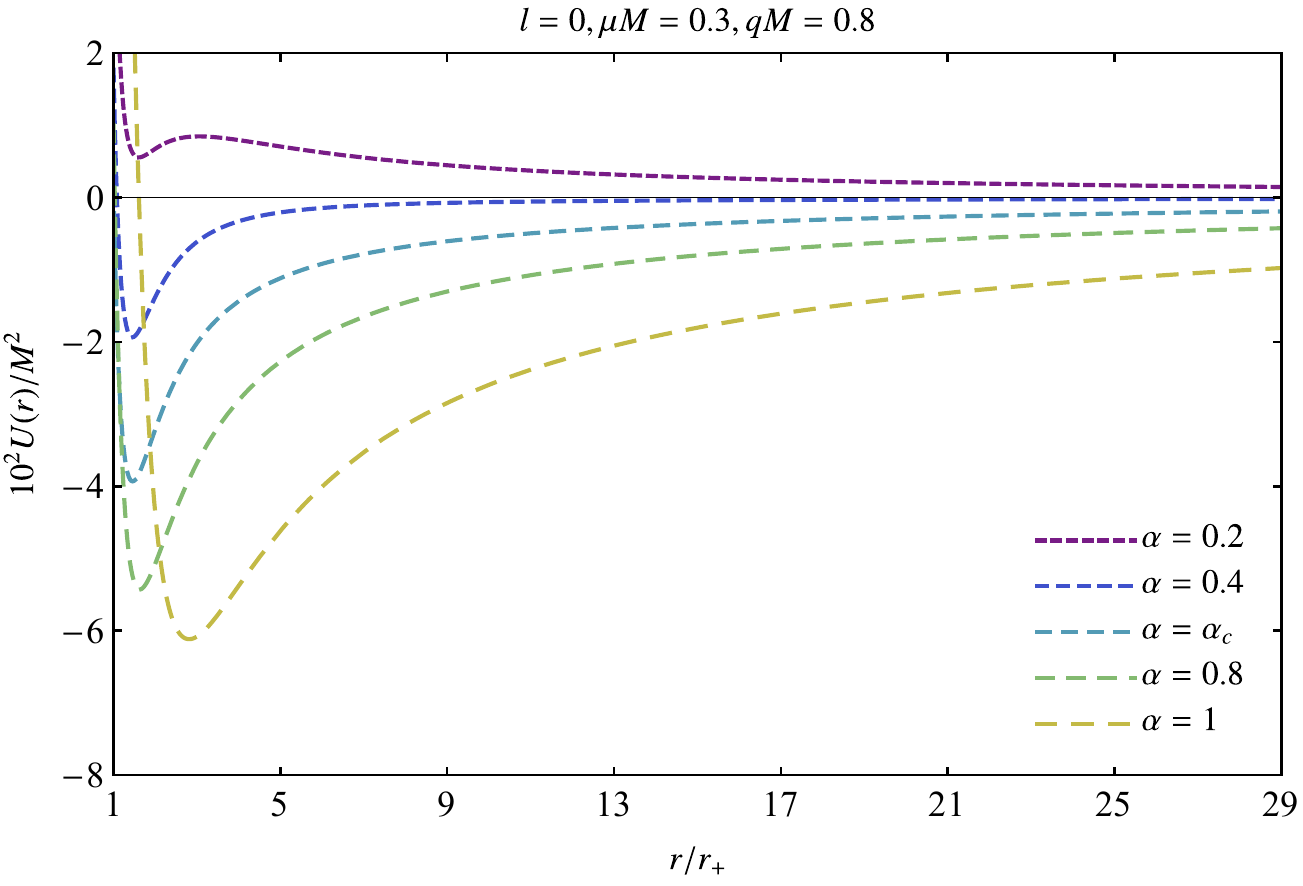}
    \caption{Effective potential $U(r)$ for ABG (top panel) and RN (bottom panel) BH spacetimes, considering $l = 0$, $\mu M = 0.3$, and $qM = 0.8$ in both cases. Moreover, $\alpha_{c} = Q_{c} / Q_{\rm{ext}} = 0.5913$ corresponds to the critical charge for the ABG and RN cases.}
    \label{effpot}
\end{centering}
\end{figure}

In Fig.~\ref{effpot}, we plot the potential given by Eq.~\eqref{newpotef} for ABG and RN BHs, considering the same parameters. We notice the existence of a propagative region from a certain radius $r_c$ out to infinity when $Q < Q_{c}$, whereas for $Q > Q_c$, the only propagative region is close to the horizon. Furthermore, although for both spacetimes we can have the propagative region, only for the ABG case we have superradiant modes ($\omega < q \phi_{+}$) satisfying $\omega^{2} > \mu^{2}$, for a given BH charge value. To illustrate this, we consider some concrete cases using the parameters presented in Fig.~\ref{effpot}.  

Notice that the absorption parameter space, which provides the regions where we have different absorption patterns in the low-frequency regime, can be mapped into a $(\mu M / q Q, Q/M)$ plot, as will be discussed later in the right panels of Fig.~\ref{apshpfunc}. Therefore, in the following examples, we also exhibit the corresponding values of  the ratio $\mu M / q Q$. When $\alpha^{\rm{ABG}} = 0.2$, $\mu = 0.3$, and $qM = 0.8$, we have $\omega_{c} \approx 0.147$ and $\mu M/qQ \approx 2.95$. In this case, we do not expect to see the unbounded superradiance phenomenon since even though we have unbounded modes, the corresponding modes are not superradiant. However, for $\alpha^{\rm{ABG}} = 0.4$, $\mu = 0.3$, and $qM = 0.8$, we have $\omega_{c} \approx 0.3016$ and $\mu M/qQ \approx 1.47$, and the unbounded superradiance phenomenon takes place, as we can check with the numerics~\cite{MAP2024b}. Hence, $Q < Q_{c}$ is not enough to guarantee the presence of unbounded superradiance in NED-sourced spacetimes. On top of that, we also need superradiant modes satisfying $\omega^{2} > \mu^{2}$.

Notice that Fig.~\ref{effpot} also shows that for $Q < Q{c}$, we have two propagative regions. One is close to the event horizon and is present even when $Q > Q_{c}$. The second extends from a given radius $r_{c}$ out to infinity. We can heuristically think that the existence of the second propagation region, given by $r \in [r_{c}, \infty)$, gives rise to a trapping region. If the trapped modes are superradiant, they can lead to the unbounded superradiance.

There is an interesting graphical way to represent the potential well associated with the unbounded superradiance by analyzing the behavior of the potential according to frequency. In order to do this, we use the method discussed in Ref.~\cite{SP2022}. We first solve $V(r) = -(\omega - \omega_{+})(\omega-\omega_{-}) = 0$ [see Eq.~\eqref{EffP}], yielding
\begin{equation}
\omega_{\pm} = q\phi(r) \pm \sqrt{f(r)\left[\mu^{2}+\dfrac{f^{\prime}(r)}{r}+\dfrac{l(l+1)}{r^{2}} \right]}.
\end{equation}
As discussed in Ref.~\cite{SP2022}, modes below $\omega_{-}$ have negative norm. These modes are associated with the superradiant scattering, as this phenomenon occurs when a wave with negative energy is absorbed by the BH. Conversely, modes above $\omega_{+}$ have positive norm.
\begin{figure}[!htbp]
\begin{centering}
    \includegraphics[width=1\columnwidth]{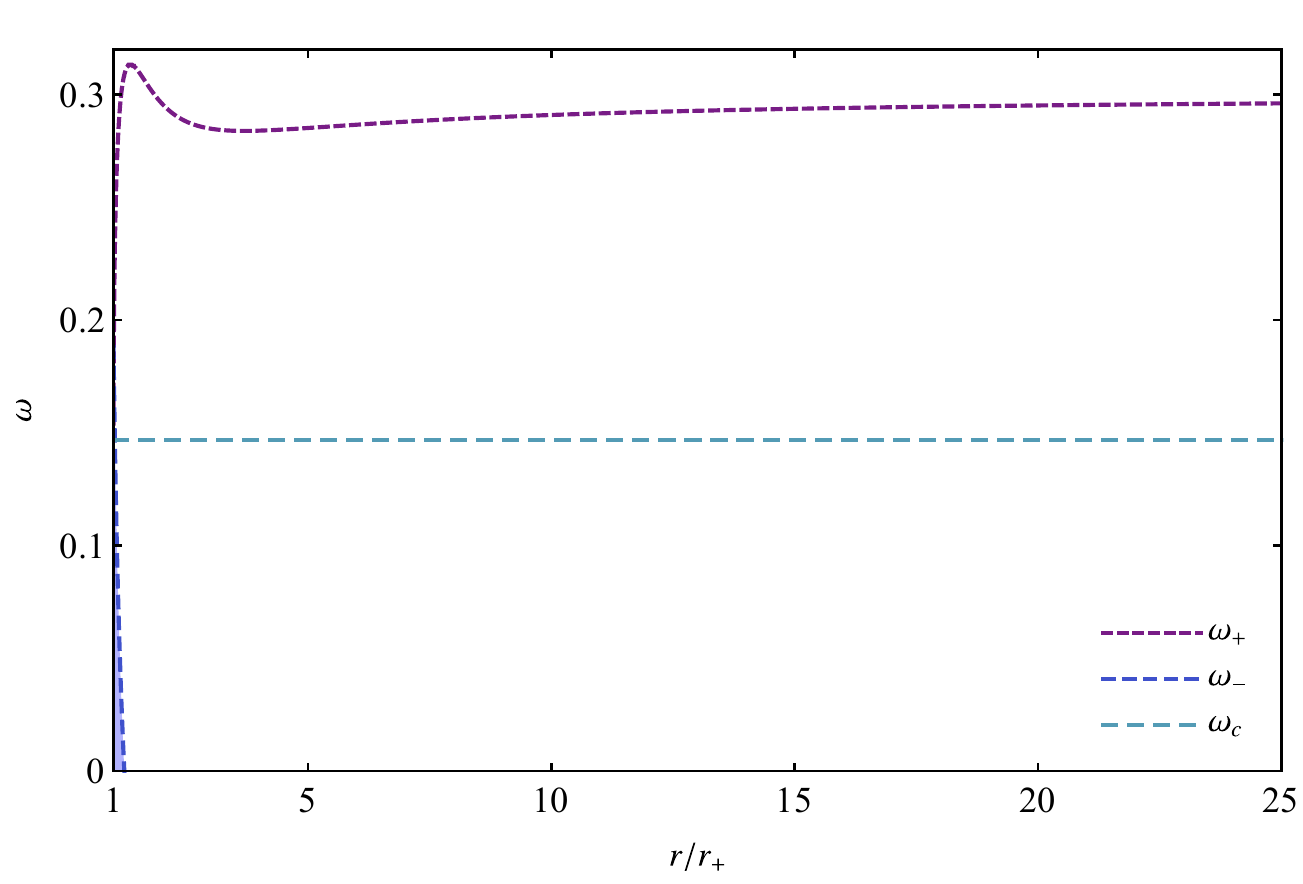}
    \includegraphics[width=1\columnwidth]{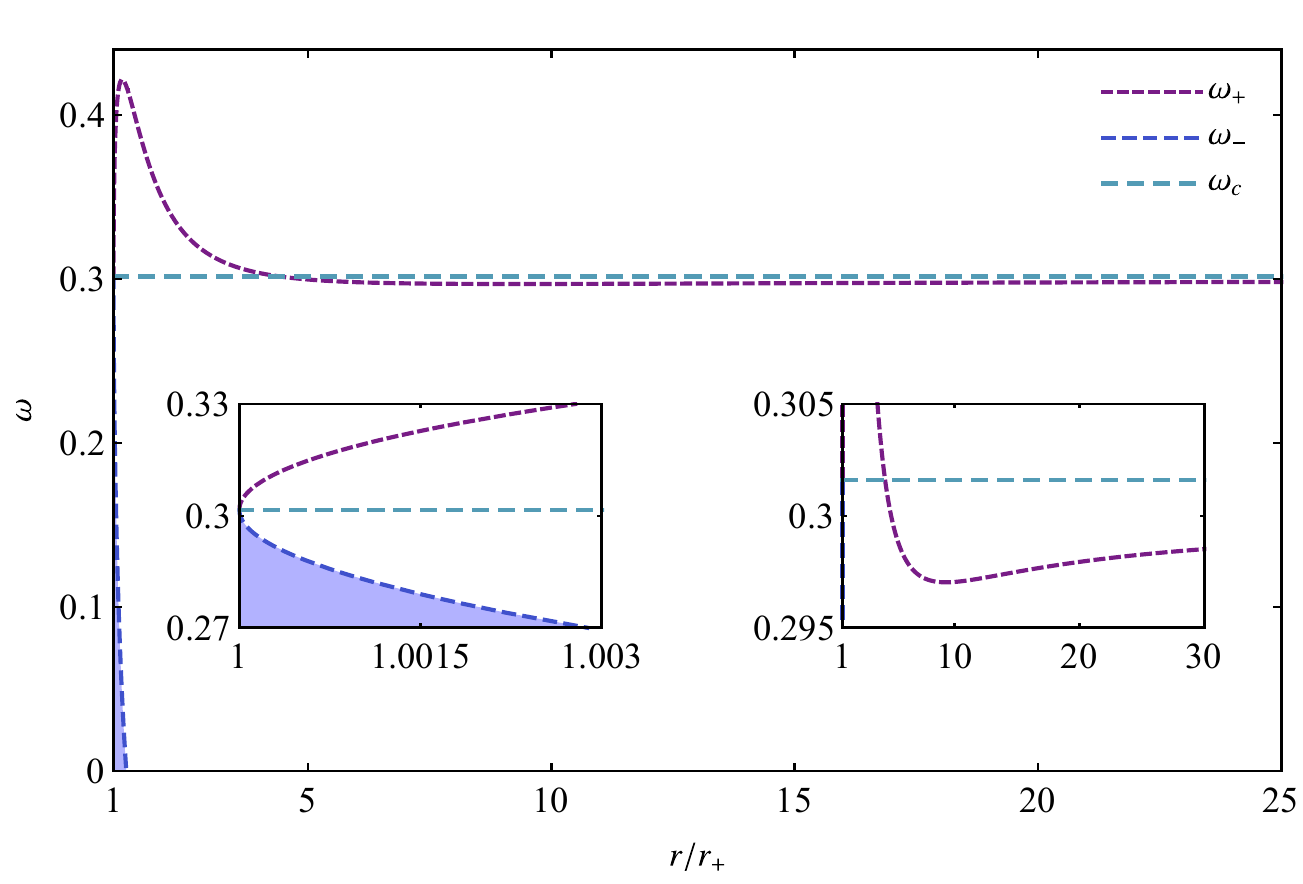}
    \includegraphics[width=1\columnwidth]{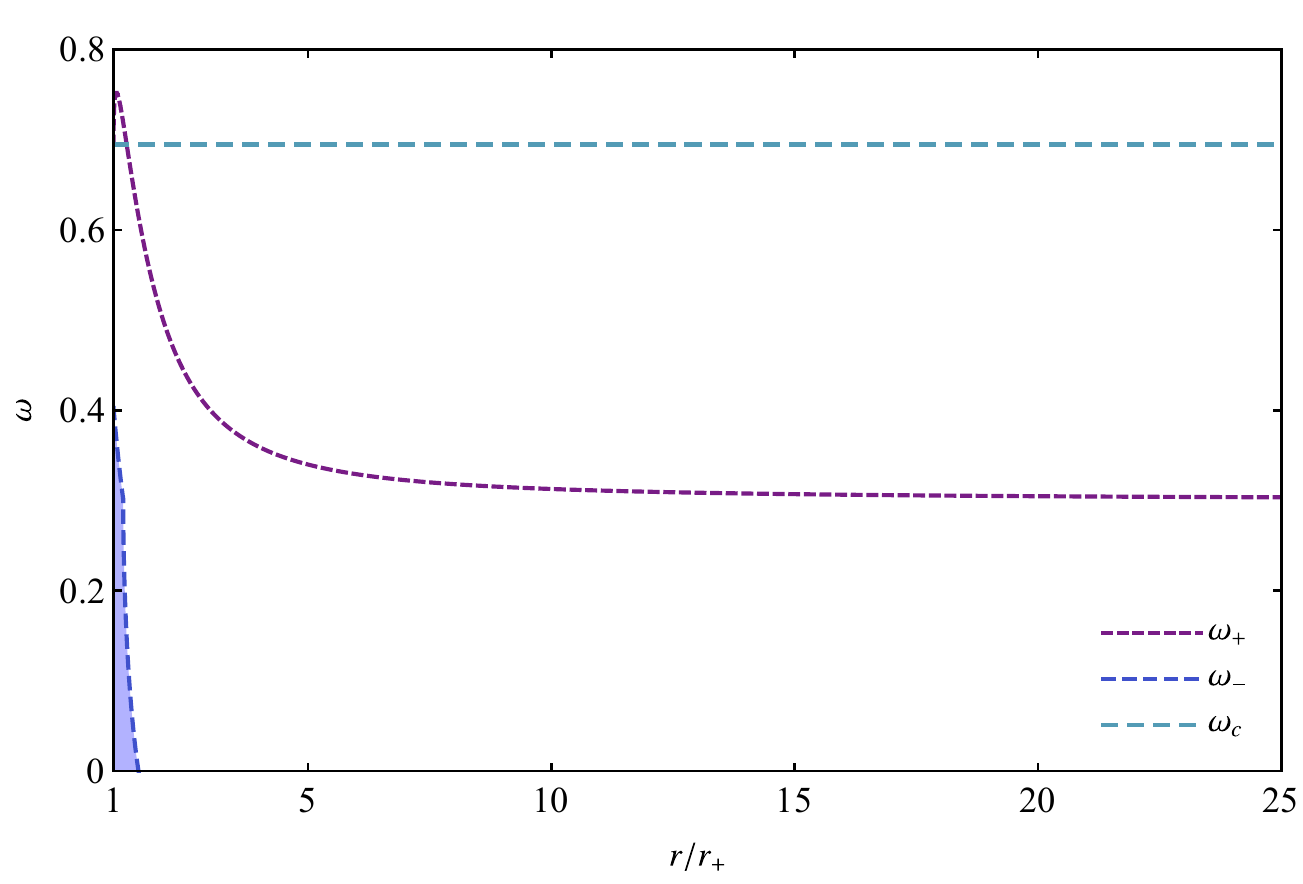}    \caption{Frequencies $\omega_{\pm}$ for $\alpha^{\rm{ABG}} = 0.2$ (top panel), $0.4$ (middle panel), and $0.8$ (bottom panel). Here, we consider $l = 0$, $\mu M = 0.3$, and $qM = 0.8$ in all panels, which are the same parameters used in Fig.~\ref{effpot}. In the middle panel, we also exhibit two insets zooming in on the frequencies near the horizon (left inset) and close to the threshold frequency (right inset) in order to improve the viewing of the trapping region. Moreover, we have $\omega_{c} = 0.147$, $0.3016$, and $0.6945$, respectively.}
    \label{frequencies}
\end{centering}
\end{figure}

In Fig.~\ref{frequencies}, we analyze the behavior of $\omega_{\pm}$ for certain configurations presented in the top panel of Fig.~\ref{effpot}. From the top to the bottom panels, we consider the situations where we have unbounded absorption, unbounded superradiance, and bounded superradiance in the ABG RBH spacetime, respectively. As we can see, in the unbounded superradiance scenario (see the middle panel of Fig.~\ref{frequencies}), we have a frequency band providing a mechanism for trapping the waves: a well located between the superradiance threshold, $\omega_c$, and the positive norm frequency threshold, $\omega_+$. This is not observed in the top and bottom panels.

Furthermore, the existence of the propagative region is closely related to the critical charge $Q_{c}$ and the Newtonian/Coulomb interactions. Recall that the large-$r$ expansion of $U(r)$ for the RN BH leads to
\begin{equation} 
\label{eq:U}U(r) = 2 \mu \frac{(\mu M - qQ)}{r} + \mathcal{O}(r^{-2}).
\end{equation}
For $\mu M > q Q$, Newtonian attraction dominates over the Coulomb repulsion and the propagative region exists. On the other hand, for $\mu M < q Q$, Coulomb repulsion is dominant and the propagative region does not exist. The critical case is at $\mu M = q Q$. Notably, the values of $Q_{c}$ correspond to the situations where $\mu M/qQ = 1$ in the absorption parameter space. The case $Q < Q_{c}$, e.g., is consistent with $\mu M > q Q$, where the Coulomb repulsion is unable to counterbalance the Newtonian attraction in order to make the cross section finite.

For NED-sourced spacetimes that behave as the RN BH for weak fields, the large-$r$ behavior of $U(r)$ coincides with that of the RN case, which is given by Eq.~\eqref{eq:U}. This expected coincidence makes it simple to find a third condition for having unbounded superradiance based on the behavior of the electrostatic potential, as we discuss below.

\subsection{Sufficient conditions}

We know that if the model reduces to the Maxwell theory in the far field, then $\mathcal{H}_{P} \rightarrow 1$ and $\mathcal{H}_{PP} \rightarrow 0$. Thus, in this limit, according to Eq.~\eqref{EP}, we have
\begin{equation}
\label{electricpotrn}\phi^{\rm{RN}}(r) = \dfrac{Q}{r},
\end{equation}
which is the radial electrostatic potential of the RN BH. In the general case, Eq.~\eqref{EP} can be written as
\begin{equation}
\label{gencase}\phi(r) = \left[\dfrac{Q \mathcal{H}_{P}}{r}\right]_{\infty}^{r}-\int_{0}^{\mathcal{P}}\left(\dfrac{-Q^{2}P}{2}\right)^{1/4}\mathcal{H}_{PP}dP,
\end{equation}
where $\mathcal{P}\equiv(-2{Q^2}/P)^{1/4}$.

The ratio, $\Phi_{\rm{hor}}$, between the radial electrostatic potentials of an NED model, $\phi^{\rm{mod}}$, and the RN BH, $\phi^{\rm{RN}}$, evaluated at the event horizon, will be defined as
\begin{equation}
\label{ratio}\Phi_{\rm{hor}} \equiv \dfrac{\phi^{\rm{mod}}(r_{+})}{\phi^{\rm{RN}}(r_{+})},
\end{equation}
where ``$\text{\rm{mod}}$'' denotes that the quantity is related to the NED model and $r_{+}$ is the event horizon location of the corresponding geometries. Then, Eq.~\eqref{ratio} leads to
\begin{align}
\nonumber \Phi_{\rm{hor}} = & \ \dfrac{Q^{\rm{mod}}}{Q^{\rm{RN}}}\dfrac{r_{+}^{\rm{RN}}}{r_{+}^{\rm{mod}}}\left(\dfrac{d \mathcal{H}}{dP} \right)\bigg|_{r = r_{+}} \\
\label{ratio2}&-\dfrac{r_{+}^{\rm{RN}}}{Q^{\rm{RN}}}\left[\int_{0}^{\mathcal{P}_{+}}\left(-\dfrac{Q^{2}P}{2}\right)^{1/4}\mathcal{H}_{PP}dP\right]_{\rm{mod}},
\end{align}
in which $\mathcal{P}_{+}$ can be obtained with the help of Eq.~\eqref{ms} evaluated at the event horizon. In the parameter space for the RN BH, the superradiance threshold exists below the line $\mu M/qQ = 1$ [see, e.g., Fig. 5 (right panel) of Ref.~\cite{MAP2024c}].

Therefore, for NED-sourced spacetimes that reduce to Maxwell's theory in the weak field limit, if we have superradiant modes satisfying $Q < Q_{c}$, we just need to enhance the strength of $\omega_{c} = q\phi_{+}$, when compared to the RN case, to make unbounded superradiant modes feasible. In other words,
\begin{equation}
\label{fundrel23}\Phi_{\rm{hor}} > 1.
\end{equation}
Thus, we claim that an unbounded superradiant regime arises in the background of electrically charged NED-sourced spacetimes (obtained in the $P$ framework) if the parameter space holds, simultaneously, the following three conditions:
\begin{subequations}
\begin{align}
\label{cond1}&\textit{(i)} \ Q < Q_{c} \text{ [see Eq.~\eqref{critcharge}]}; \\
\label{cond2}&\textit{(ii)} \ \omega < q\phi(r_{+}) \text{ [see Eq.~\eqref{criticalfreq}]}; \\
\label{cond3}&\textit{(iii)} \ \phi^{\rm{mod}}(r_{+}) > \phi^{\rm{RN}}(r_{+}) \text{ [see Eq.~\eqref{fundrel23}]}.
\end{align}
\end{subequations}
The condition (i) guarantees the existence of the unbounded regime, and (ii) is necessary for the superradiant scattering. Last but not least, (iii) makes it feasible for superradiant scalar wave modes to exist in the region where we have an unbounded absorption, leading to unbounded superradiance.

\begin{figure*}[!htbp]
\begin{centering}
    \includegraphics[width=1\columnwidth]{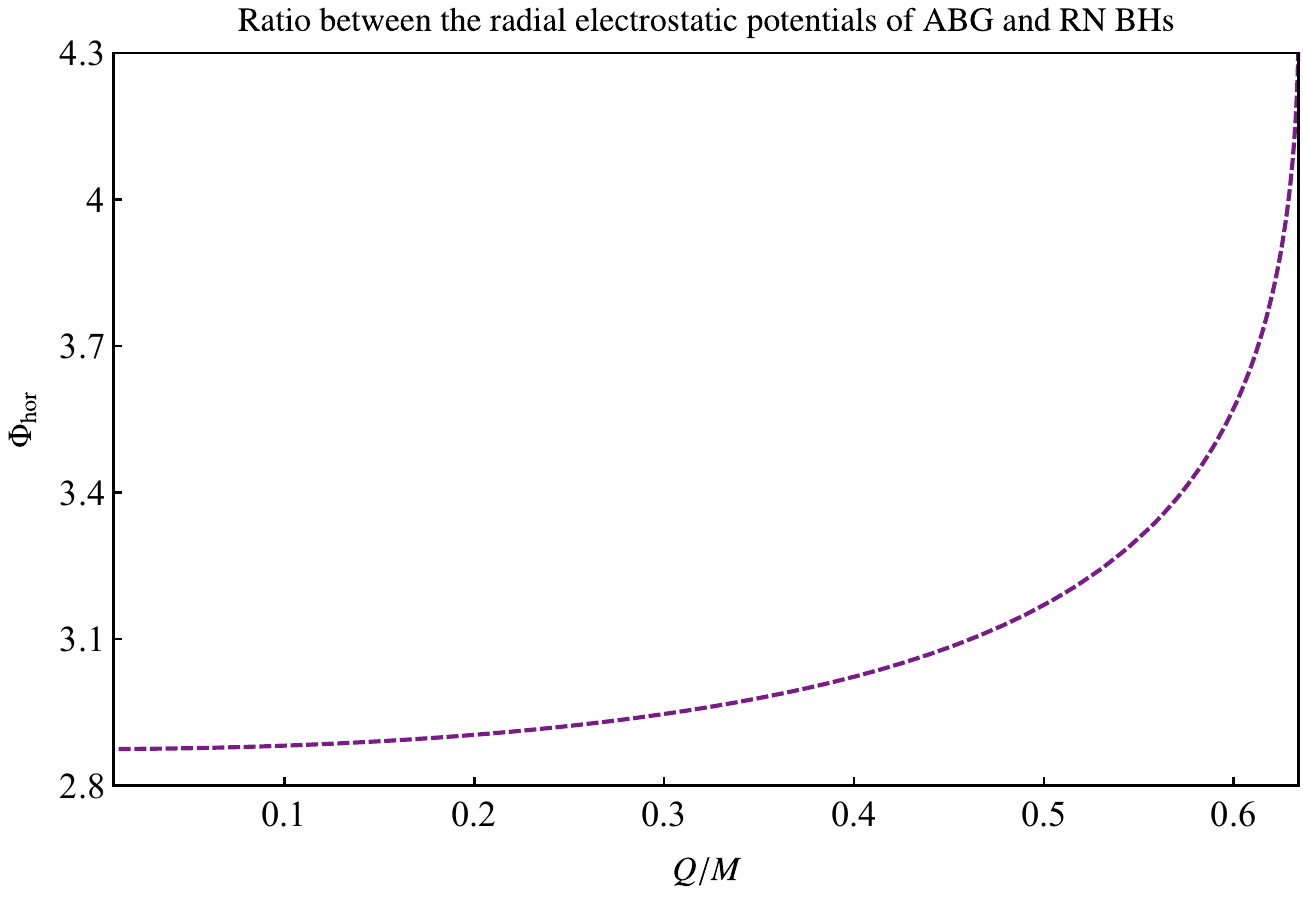}
    \includegraphics[width=1\columnwidth]{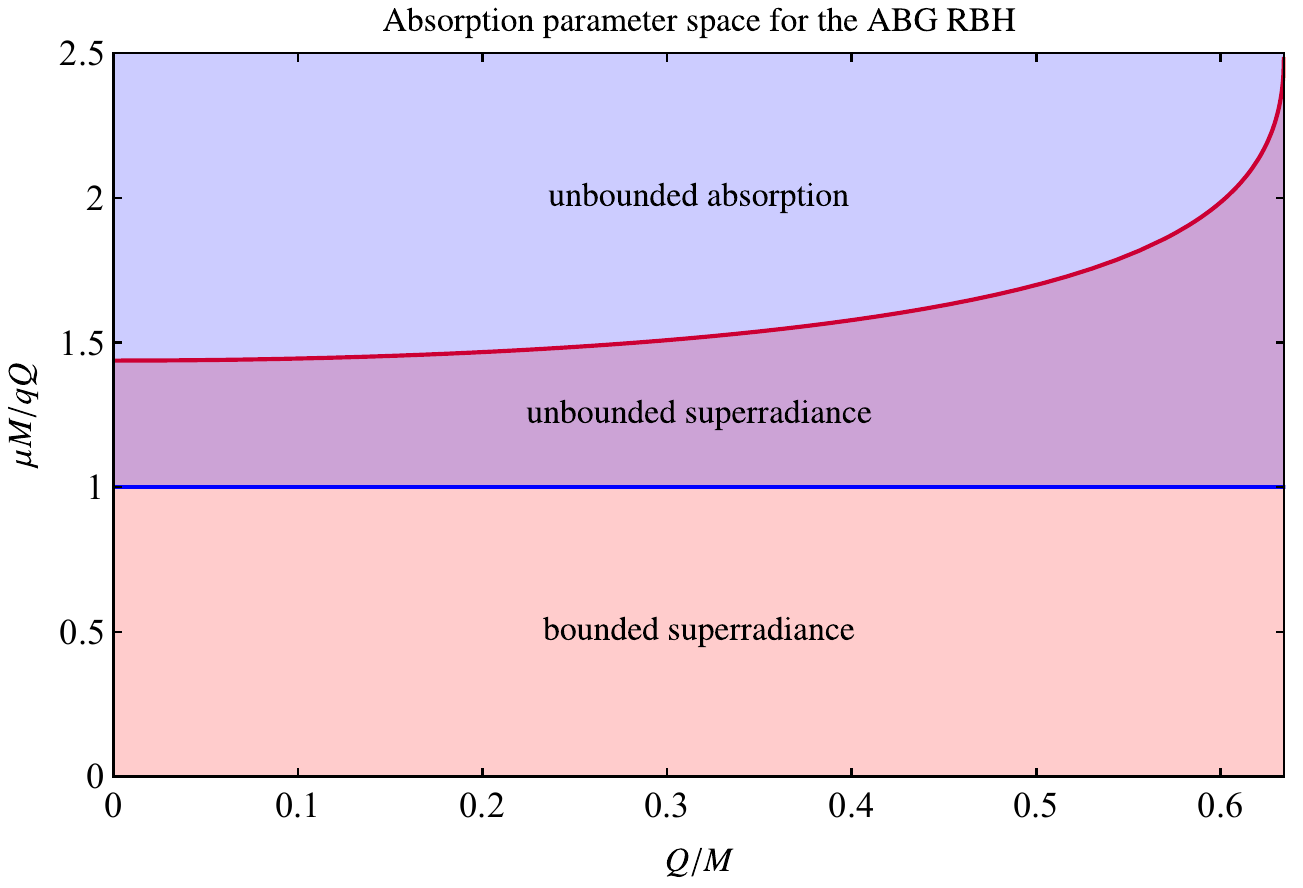}
    \includegraphics[width=1\columnwidth]{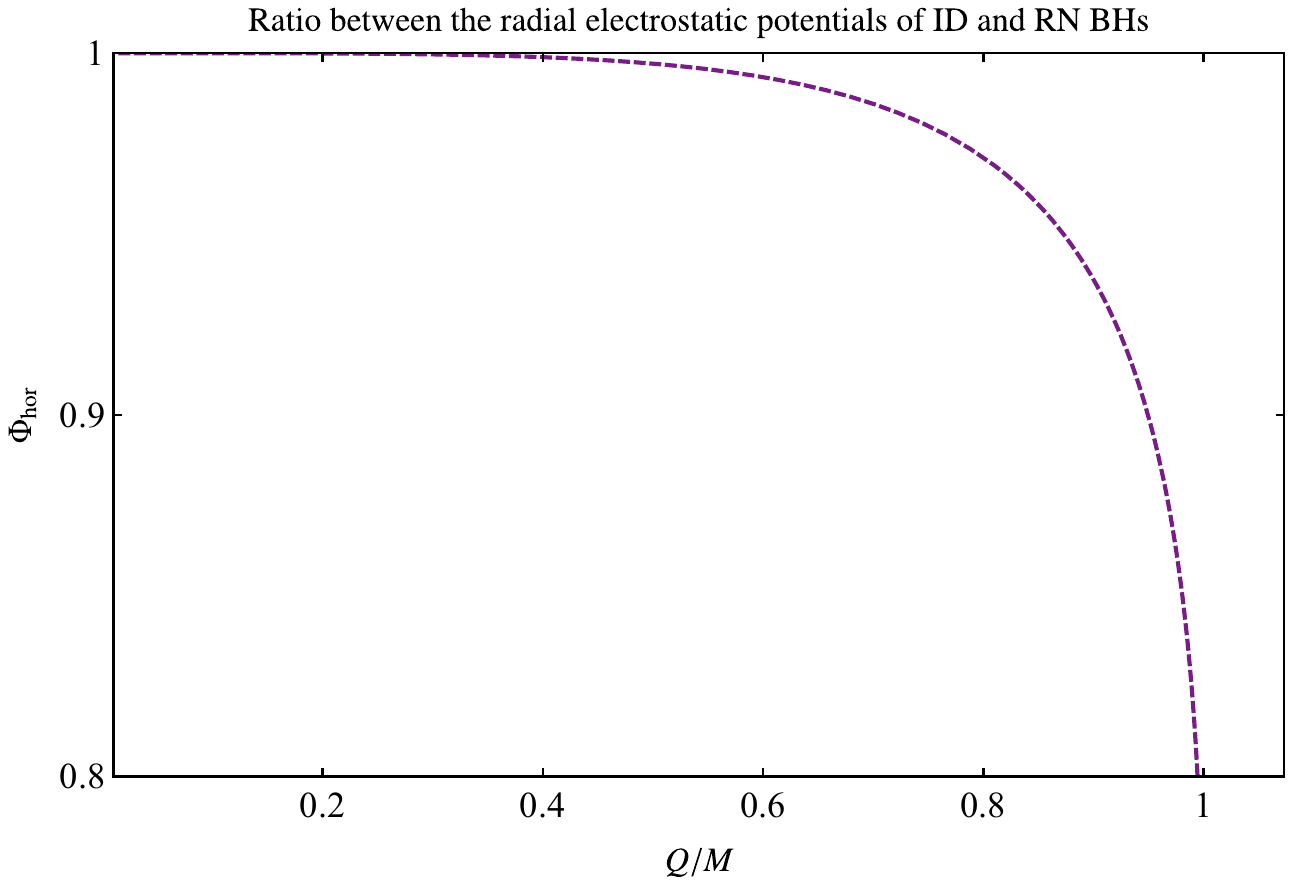}    
    \includegraphics[width=1\columnwidth]{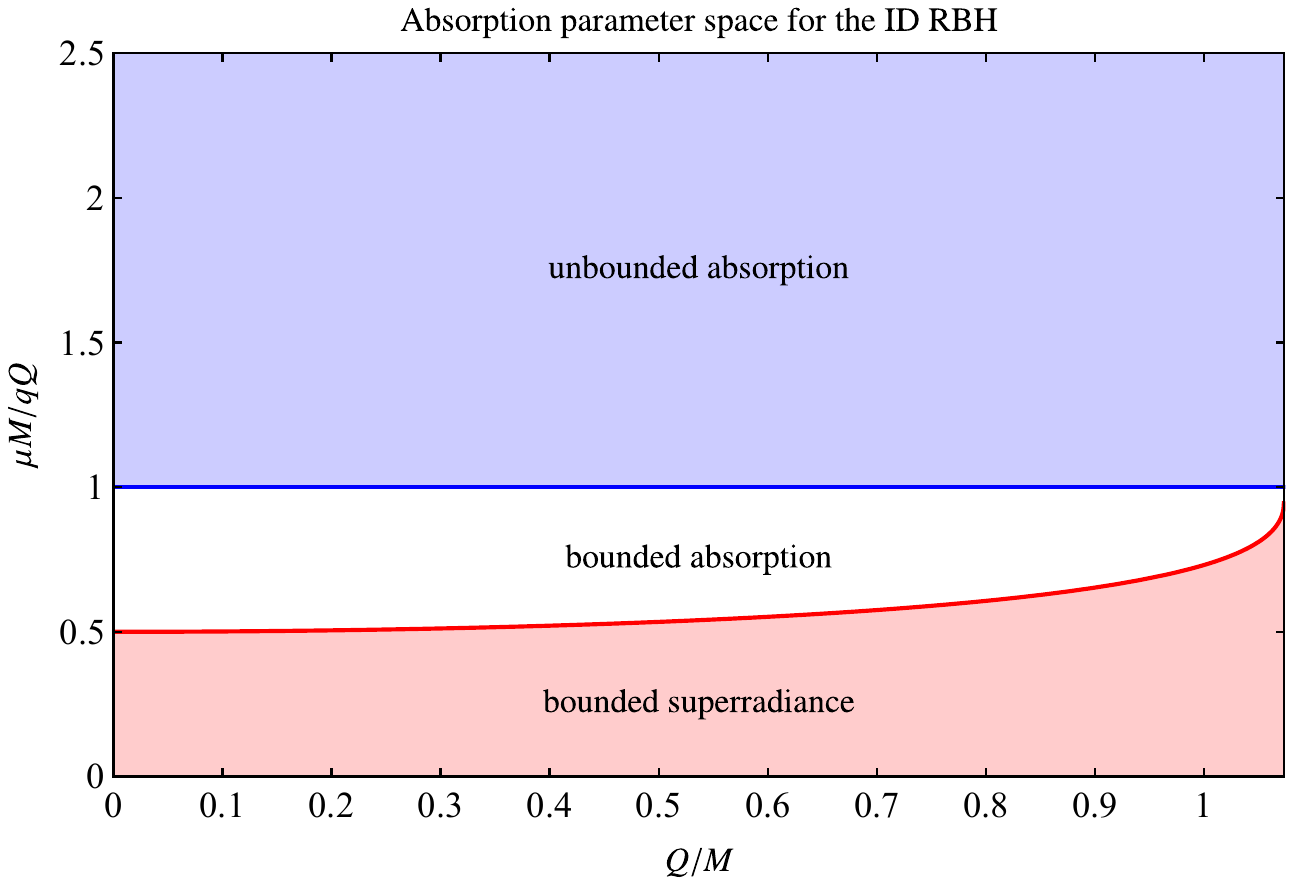}
    \includegraphics[width=1\columnwidth]{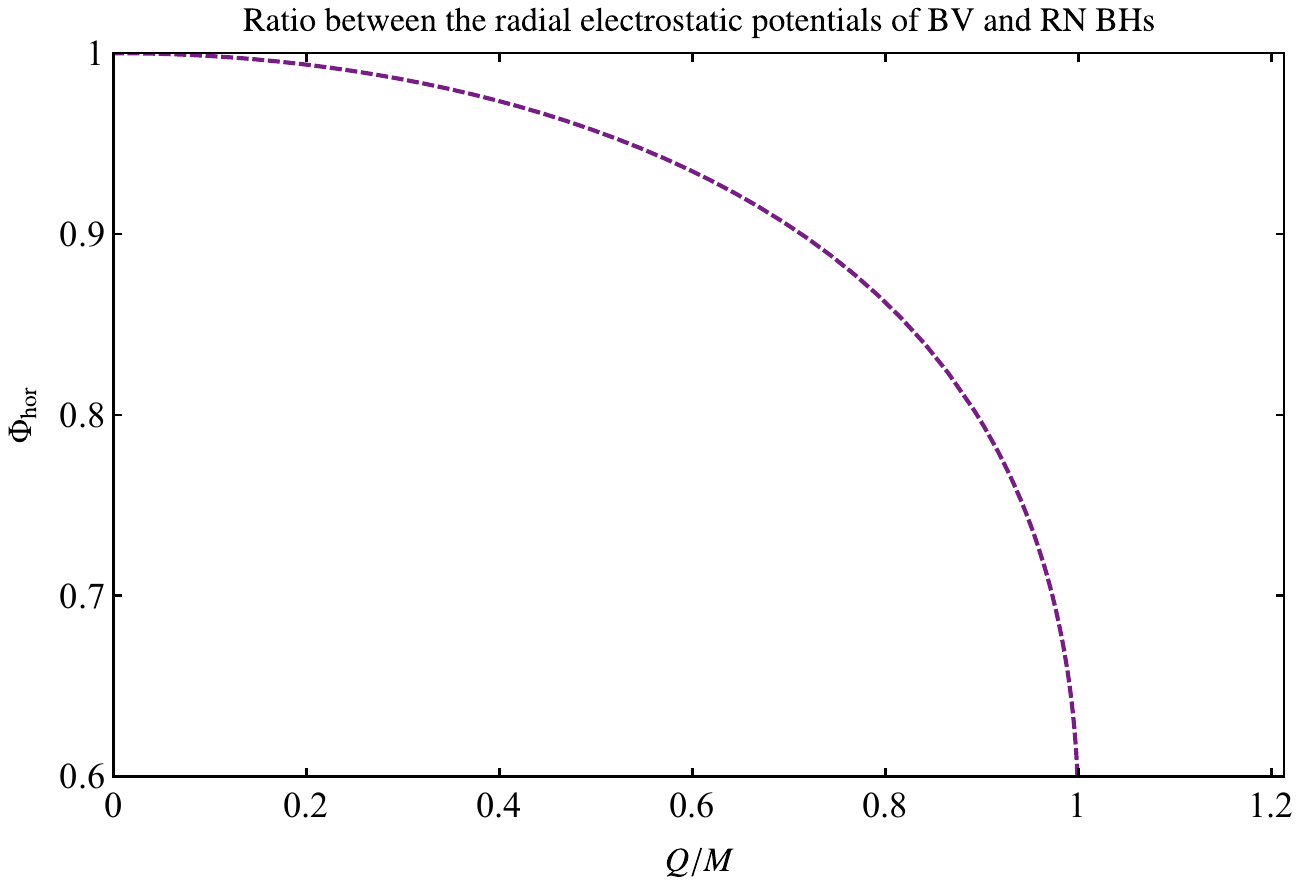}
    \includegraphics[width=1\columnwidth]{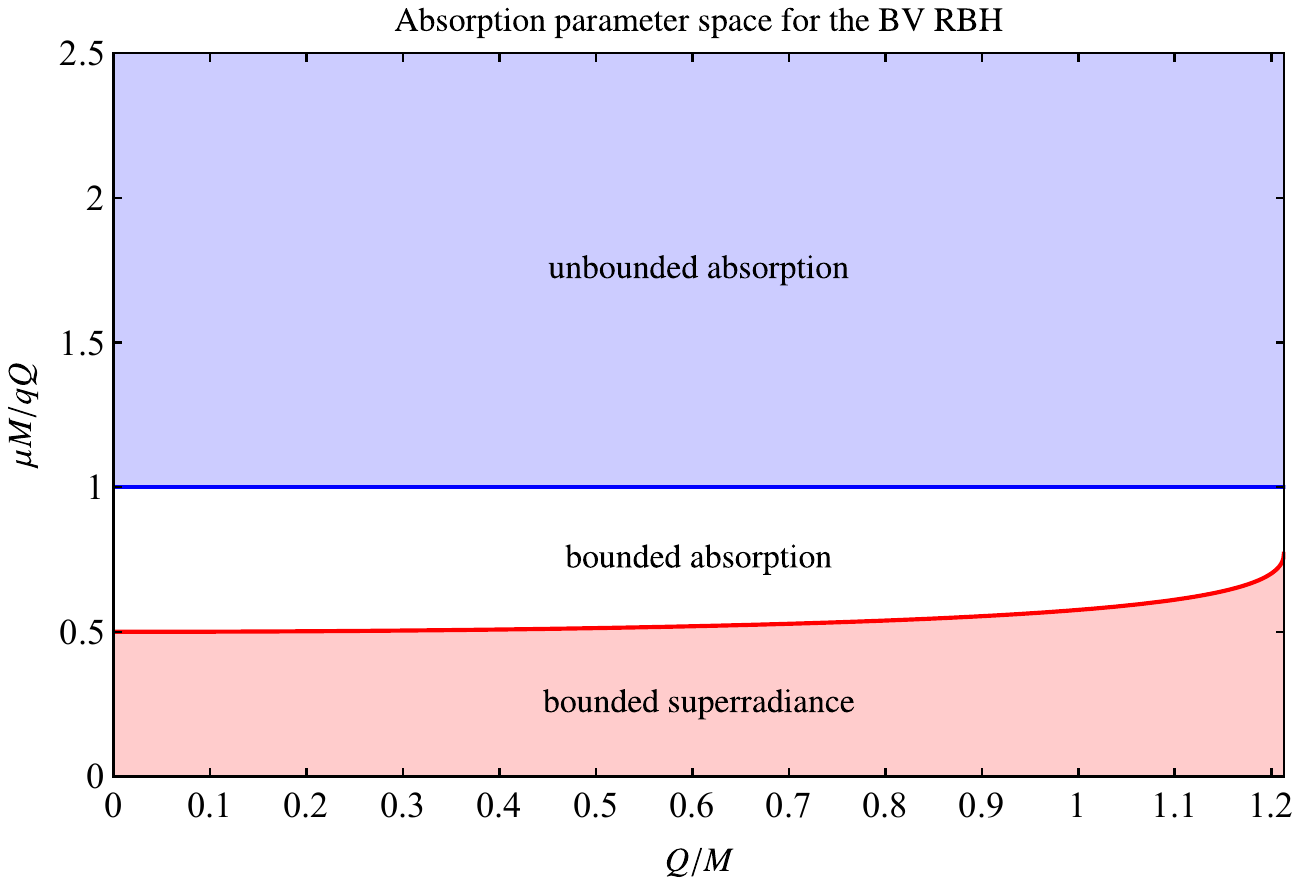}
    \caption{Ratio between the radial electrostatic potential of some well-known NED-based electrically charged RBHs and that of the RN BH (left panels), as functions of $Q/M$, and the absorption parameter space of such RBH geometries (right panels), considering a charged massive scalar wave. In the right panels, the solid red curve corresponds to the superradiance threshold and the solid blue line to the absorption parameter space that satisfy Eq.~\eqref{critcharge}. From the top to the bottom, we consider ABG~\cite{ABG1998}, ID~\cite{D2004}, and BV models~\cite{BV2014}. We observe that the unbounded superradiance occurs only in the first RBH model~\cite{ABG1998} (top panel), in agreement with our sufficient conditions.}
    \label{apshpfunc}
\end{centering}
\end{figure*}

In Fig.~\ref{apshpfunc}, we show the ratio between the electrostatic potential of some well-known electrically charged RBH spacetimes evaluated at the event horizon, considering distinct values of the BH charge-to-mass ratio, and their corresponding absorption parameter space. We observe the existence of an unbounded superradiant regime only in the ABG RBH geometry. For this solution, all three conditions mentioned earlier are satisfied. However, the other models present an absorption parameter space similar to the RN case, i.e., they have unbounded absorption and bounded absorption with and without superradiance --- but no unbounded superradiant regime. Therefore, for these models, as in the RN case, it is not possible to find configurations satisfying the sufficient conditions to have unbounded superradiance. 

\subsection{Exploring the electrostatic potentials}

We want to obtain a condition for which the superradiance threshold [see, e.g., Eq.~\eqref{criticalfreq}] for the NED model is higher than that in the RN case, for the same field charge values. In other words, we want to see in which possible scenarios we have $\Phi_{\rm{hor}} > 1$. To do this, let us now break down Eq.~\eqref{ratio2} term by term. At the event horizon, if we require that $\mathcal{H}_{P} > 1$ (the Maxwell case is given by $\mathcal{H}_{P} = 1$), then the first term on the right-hand side of Eq.~\eqref{ratio2} is higher than unity if, for the same BH charges, $r_{+}^{\rm{RN}} > r_{+}^{\rm{mod}}$. The behavior of the second term is subtler. We cannot perform the integration without knowing the NED model. However, if the integrand of the second term is negative in the domain $P = [0,\mathcal{P}_{+}]$ (or, equivalently, $r = [\infty,r_{+}]$), then the contribution of the second term increases $\Phi_{\rm{hor}}$. This possibility can only be achieved if $\mathcal{H}_{PP} < 0$, since $(-Q^{2}P/2)^{1/4} > 0$, for electrically charged BHs in NED [cf. Eq.~\eqref{ms}]. 
\begin{figure*}[!htbp]
  \centering
 \subfigure[\ ]{\includegraphics[width=1\columnwidth]{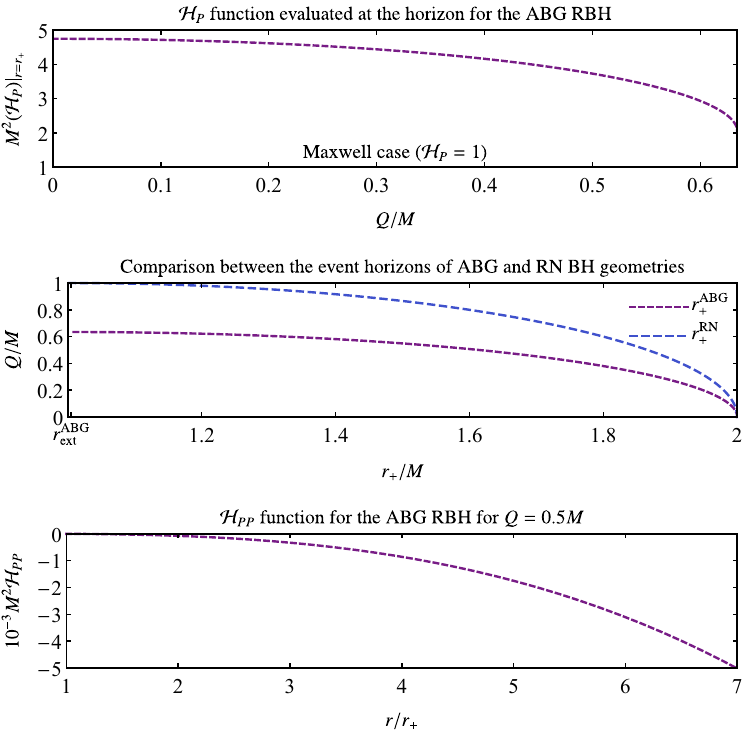}\label{a}} \hspace{0.2cm}
  \subfigure[\ ]{\includegraphics[width=1\columnwidth]{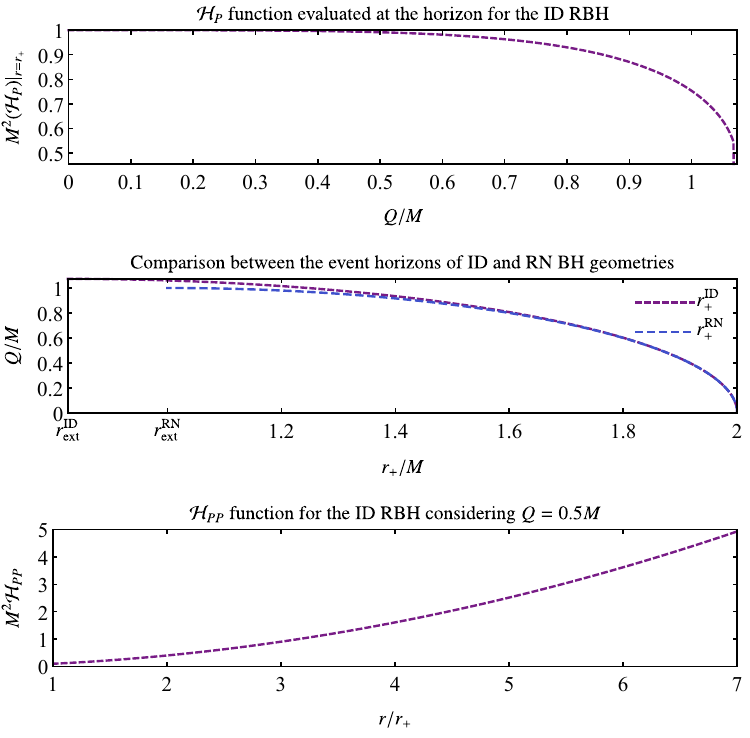}\label{c}} \hspace{0.2cm}
    \subfigure[\ ]{\includegraphics[width=1\columnwidth]{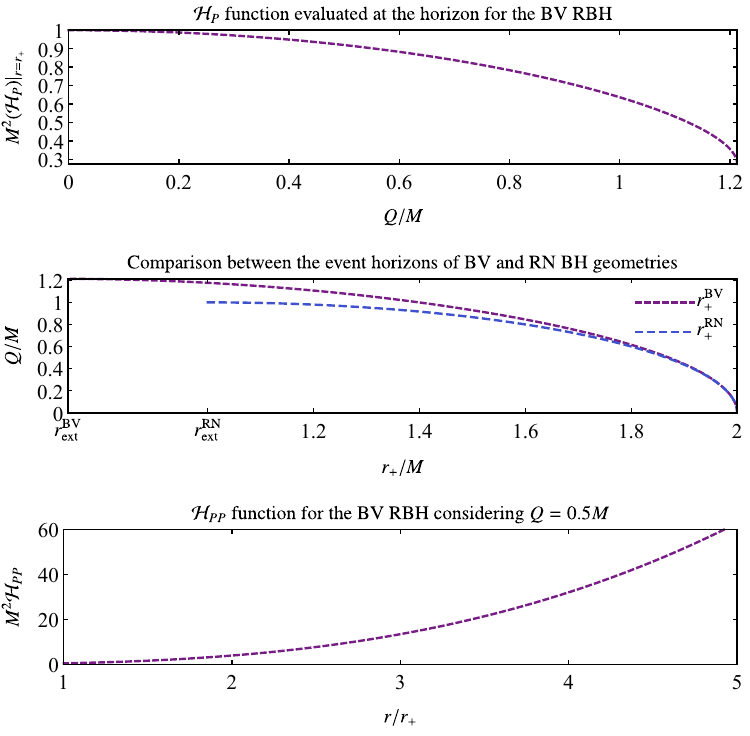}\label{d}}
\caption{Conditions for $\phi^{\rm{mod}}(r_{+}) > \phi^{\rm{RN}}(r_{+})$, considering some well-known RBH solutions. (a) Conditions for $\phi^{\rm{ABG}}(r_{+}) > \phi^{\rm{RN}}(r_{+})$, considering the ABG solution~\cite{ABG1998}. Here, $r_{\rm{ext}}^{\rm{ABG}} \approx 1.005M$. (b) Conditions for $\phi^{\rm{ID}}(r_{+}) > \phi^{\rm{RN}}(r_{+})$, considering the ID model~\cite{D2004}. Here, $r_{\rm{ext}}^{\rm{ID}} \approx 0.825M$. (c) Conditions for $\phi^{\rm{BV}}(r_{+}) > \phi^{\rm{RN}}(r_{+})$, considering one of the BV models derived in Ref.~\cite{BV2014}. Here, $r_{\rm{ext}}^{\rm{ID}} \approx 0.735M$.}
\label{horchahpp}
\end{figure*} 

Under these circumstances, we obtain that 
\begin{equation}
\label{fundrel2}\phi^{\rm{mod}}(r_{+}) > \phi^{\rm{RN}}(r_{+}),
\end{equation}
which is consistent with Eq.~\eqref{fundrel23}, and, therefore, the superradiance threshold for the scalar wave in the background of the NED-based BH spacetime is higher than that in the RN case, considering fixed field charges. We emphasize that it is possible to argue that other situations lead to $\Phi_{\rm{hor}} > 1$. For example, if $\mathcal{H}_{PP}$ is positive but not enough to make $\Phi_{\rm{hor}} < 1$, then Eq.~\eqref{fundrel2} still holds. However, as we can anticipate from Fig.~\ref{horchahpp}, the conditions we have imposed for having $\Phi_{\rm{hor}} > 1$ are consistent with some well-known NED models.

In Fig.~\ref{horchahpp}, we explore the validity of the general conditions for having $\phi^{\rm{mod}}(r_{+}) > \phi^{\rm{RN}}(r_{+})$, considering some well-known electrically charged RBH spacetimes. As we can see, for the spacetimes considered, only the ABG RBH satisfies the necessary requirements for having $\phi^{\rm{mod}}(r_{+}) > \phi^{\rm{RN}}(r_{+})$. For the other models, we typically observe that $\mathcal{H}_{P}|_{r = r_{+}} < 1$, $r_{+}^{\rm{RN}} < r_{+}^{\rm{mod}}$, considering $Q^{\rm{RN}} = Q^{\rm{mod}}$, and $\mathcal{H}_{PP} > 0$. Therefore, for the models treated here, only the ABG RBH solution can provide the necessary conditions to have unbounded superradiance. Either way, the key ingredient is that Eq.~\eqref{fundrel2} is satisfied, regardless of the specific conditions satisfied by the electric NED source itself. Notice that, to avoid misunderstanding, we refer to the first ABG solution obtained in 1998 as ABG RBH spacetime. Moreover, we consider the BV exponential distribution as the NED source.

\subsection{Absorption cross section}

\begin{figure}[!htbp]
\begin{centering}
    \includegraphics[width=1\columnwidth]{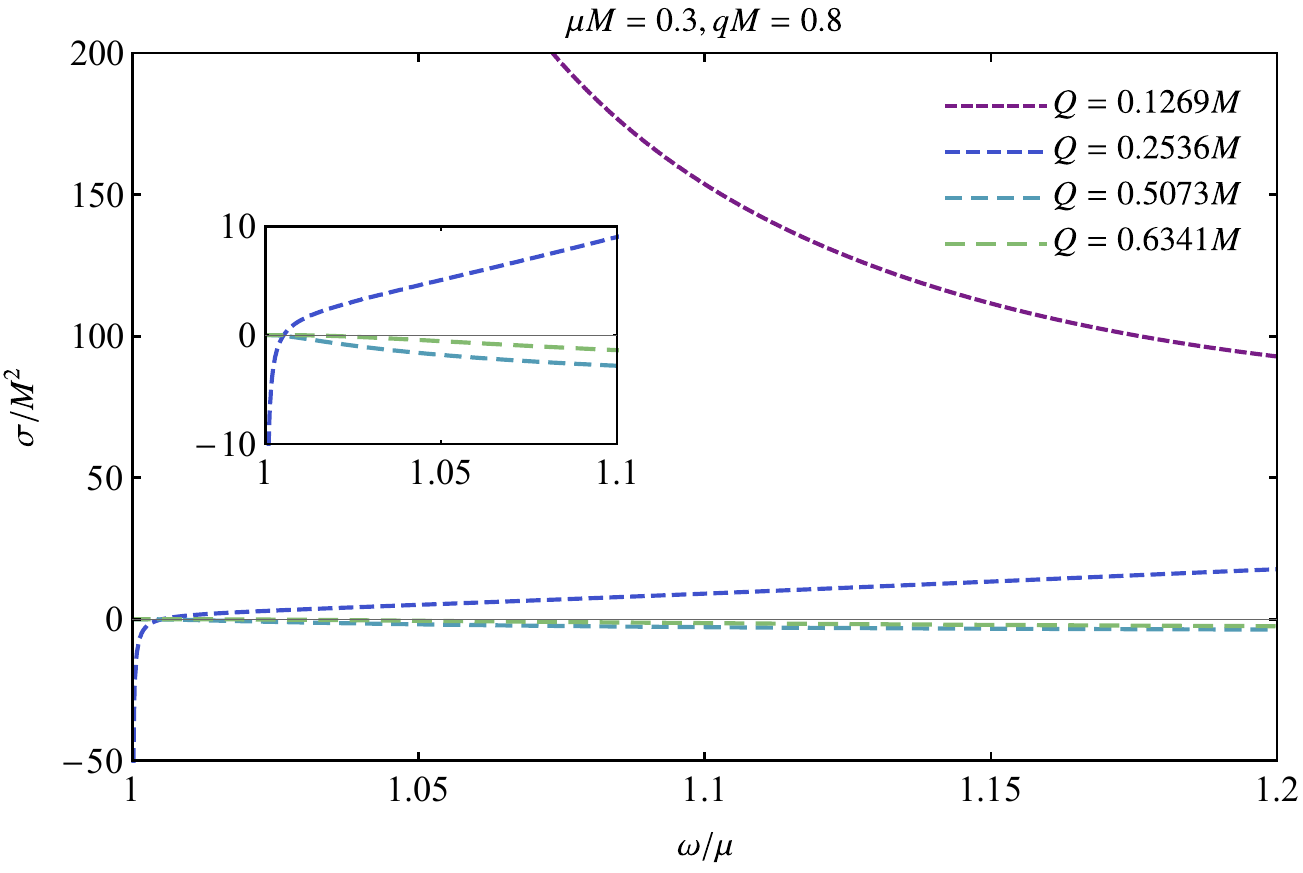}
    \includegraphics[width=1\columnwidth]{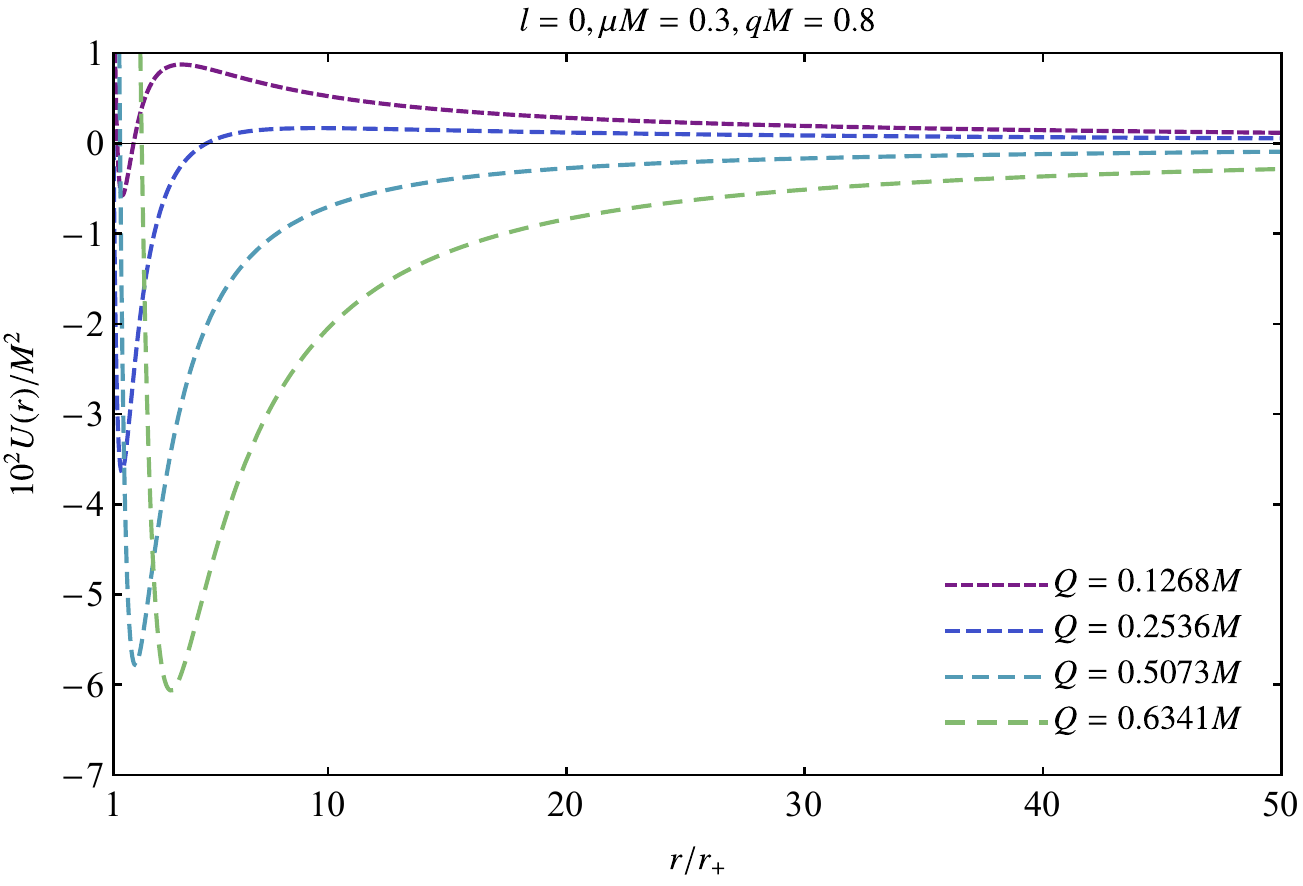}
    \caption{Top panel: total absorption cross section of the ABG RBH in the low-frequency regime, as a function of $\omega/ \mu$, for $\mu M = 0.3$, $qM = 0.8$, and distinct values of $Q/M$. Bottom panel: effective potential $U(r)$ for the ABG RBH spacetime, considering $l = 0$, $\mu M = 0.3$, $qM = 0.8$, and different choices of $Q/M$. In this setup, the critical charge is given by $Q_{c} \approx 0.375M$. For $Q = 0.2536$, we have unbounded superradiance when $1.0053 \gtrsim \omega/\mu > 1$.}
    \label{acsfiniteness}
\end{centering}
\end{figure}

In Fig.~\ref{acsfiniteness}, we explore the relationship  between the effective potential, particularly the condition given by Eq.~\eqref{critcharge}, and the total absorption cross section. In the top panel of Fig.~\ref{acsfiniteness}, we exhibit the total absorption cross section, as a function of the ratio $\omega/\mu$, for small values of the frequency. The inset helps us to observe the behavior of the cross section for $\omega/\mu\rightarrow 1$. For its turn, in the bottom panel of Fig.~\ref{acsfiniteness}, we show the potential function, considering distinct values of the BH's charge-to-mass ratio, $Q/M$. For simplicity, we consider the absorption parameter space of the ABG RBH spacetime only. As shown, for $Q < Q_{c}$, we observe the unbounded regime (e.g., the curved associated with $Q = 0.1268M$). Conversely, for $Q > Q_{c}$, we have the bounded regime (e.g., the curve associated with $Q = 0.5073M$). In Fig.~\ref{acsfiniteness}, $Q_{c} \approx 0.375M$.

It is important to note that there are two scenarios where $Q < Q_{c}$. However, only the case $Q = 0.2536M$ leads to the unbounded superradiance. This can be understood by noting that for $Q = 0.1268M$, superradiant scattering occurs only for $0 < \omega M\lesssim 0.147$. On the other hand, for $Q = 0.2536M$, superradiant scattering occurs for $0 < \omega M \lesssim 0.3016$. Therefore, only in scenarios with $Q = 0.2536M$, the frequency satisfies the proper range for superradiance and unbounded modes, i.e., $\omega_{c} > \omega > \mu$ (namely, $0.3016 > \omega M > 0.3$, for this configuration). Thus, the condition~\eqref{cond2} is fulfilled. For $Q > Q_{c}$, the absorption spectrum is related to the bounded superradiance.

The results presented in Fig.~\ref{acsfiniteness} indicate that there is a relationship between the critical charge $Q_{c}$ and the existence of the limit $\lim_{\omega \rightarrow \mu}|\mathcal{T}_{\omega l}|^{2}/\kappa^{3}$, which is associated with our definition of bounded/unbounded absorption. As we have noticed, for $Q < Q_{c}$, the total absorption cross section diverges, and, consequently, the limit does not exist. On the other hand, for $Q > Q_{c}$, the cross section is finite, and the limit exists.

\section{Concluding remarks}

In recent years, NED sources have been widely used to obtain RBH solutions, and the study of general results for NED models in BH physics has been often addressed in the literature. We have aimed to contribute to a better understanding of the applications and implications of NED models in BH physics. We have obtained sufficient conditions for having an unbounded superradiant regime in NED-based electrically charged RBH geometries: (i) $Q < Q_{c}$; (ii) $\omega_{c} < q\phi(r_{+})$; and (iii) $\phi^{\rm{mod}}(r_{+}) > \phi^{\rm{RN}}(r_{+})$ [cf. Eqs.~\eqref{cond1}-\eqref{cond3}].

Although these results are valid for a broad family of electrically charged BH solutions derived in the setup described in Sec.~\ref{sec:frame}, we have sought for more specific conditions for the electric NED sources themselves. In this sense, we have shown that for electrically charged BHs derived in the $P$ framework, a situation where $(\mathcal{H}_{P})|_{r = r_{+}} > 1$, with $r_{+}^{\rm{mod}} < r_{+}^{\rm{RN}}$, for $Q^{\rm{RN}} = Q^{\rm{mod}}$, and $\mathcal{H}_{PP} < 0$ usually leads to $\phi^{\rm{mod}} > \phi^{\rm{RN}}$, and, therefore, satisfies the third condition~\eqref{cond3}. For electrically charged BHs derived in the $F$ framework, obtaining more specific conditions for the electric NED sources is not straightforward, due to the complicated form of the function $\Phi_{\rm{hor}}$ in this case [cf. Eq.~\eqref{ratio3}].

We also notice that although there is a connection between the concepts of unbounded superradiance in the absorption cross section and the existence of superradiant bound states in the quasibound state spectrum, these are distinct ideas. Thus, we stress out that our work is devoted to the sufficient conditions for the existence of an unbounded superradiance regime in spacetimes sourced by NED. Therefore, it does not necessarily imply the phenomena of superradiant instability. We aim to address possible generalizations in this context elsewhere.

In a certain sense, this work complements the results obtained in Ref.~\cite{MAP2024c} by answering the following question: Is it possible to find sufficient conditions for having an unbounded superradiant regime in the parameter space of NED-based electrically charged BHs? We have found that the answer is \textit{yes} [see Eqs.~\eqref{cond1}-\eqref{cond3}]. We also point out that here we have considered only spacetimes sourced by NED models that correspond to linear electrodynamics in the weak field limit.

A possible extension of this work is to investigate the absorption parameter space of higher-order spin fields, e.g., electromagnetic and gravitational fields, taking into account the contributions of the nonlinear electromagnetic field. In addition, it would also be interesting to address how the stability of the spacetimes sourced by NED models is related to the sufficient conditions for unbounded superradiance.

We end this paper by discussing possible extensions beyond NED models. The condition given by Eq.~\eqref{fundrel23} is related solely to the superradiance threshold, i.e., $q\phi_{+}$. Here, we have considered that the source of the electrostatic potential is an NED model that behaves as linear electrodynamics for weak fields. Nevertheless, if we have alternative sources with Maxwell's asymptotics that result in a Klein-Gordon equation with the form given by Eq.~\eqref{KG}, we expect that the conditions presented here will potentially hold.

\begin{acknowledgments}

The authors would like to thank Sam Dolan for his insightful comments and discussions concerning this paper. The authors also thank Funda\c{c}\~ao Amaz\^onia de Amparo a Estudos e Pesquisas (FAPESPA),  Conselho Nacional de Desenvolvimento Cient\'ifico e Tecnol\'ogico (CNPq) and Coordena\c{c}\~ao de Aperfei\c{c}oamento de Pessoal de N\'{\i}vel Superior (Capes) -- Finance Code 001, in Brazil, for partial financial support. This work has further been supported by the European Union's Horizon 2020 research and innovation (RISE) programme H2020-MSCA-RISE-2017 Grant No. FunFiCO-777740 and by the European Horizon Europe staff exchange (SE) programme HORIZON-MSCA-2021-SE-01 Grant No. NewFunFiCO-101086251. We thank the anonymous referee for the careful reading and for valuable comments and suggestions on the paper.

\end{acknowledgments}

\section*{Data Availability}

No data were created or analyzed beyond what is included in this article.

\begin{appendix}

\section{ELECTRICALLY CHARGED BLACK HOLES IN NONLINEAR ELECTRODYNAMICS}\label{apx}

Throughout this work, we consider some BH solutions obtained in NED as sample tests of our main findings. Here, we present the corresponding structural function $\mathcal{H}(P)$, $f(r)$, and $\phi(r)$ associated with these models, which are the following

\begin{itemize}
\item First ABG model~\cite{ABG1998}: 
\begin{align}
\mathcal{H}(P) & = P \dfrac{1-3 \mathcal{Y}}{\left(1+\mathcal{Y}\right)^{3}}-\dfrac{6}{s Q^{2}}\left(\dfrac{\mathcal{Y}}{1+\mathcal{Y}} \right)^{5/2}, \\
f(r) & = 1-\frac{2 M r^2}{\left(Q^2+r^2\right)^{3/2}}+\frac{Q^2 r^2}{\left(Q^2+r^2\right)^2}, \\
\phi(r) & = \dfrac{3M}{2Q}+\dfrac{1}{2}Qr^{5}\left(\dfrac{2}{(Q^{2}+r^{2})^{3}}-\dfrac{3M}{Q^{2}(Q^{2}+r^{2})^{5/2}}\right).
\end{align}
\item ID model~\cite{D2004}: 
\begin{align}
\mathcal{H}(P) & = \dfrac{P}{\left(1+\alpha \sqrt{-P}\right)^{2}}, \\
f(r) & = 1 - \dfrac{4M}{\pi r}\left[\arctan\left(\dfrac{r}{z}\right)-\dfrac{rz}{r^{2}+z^{2}} \right], \\
\nonumber\phi(r) & =\dfrac{1}{2 \sqrt{2 \pi }}\bigg[\frac{\sqrt{M z} \left(5 r^3+3 r z^2\right)}{\left(r^2+z^2\right)^2}+\\
& \ \quad \frac{3 M }{2 \sqrt{M z}}	\left(\pi -2 \tan^{-1}\left(\frac{r}{z}\right)\right)\bigg].
\end{align}
\item BV model (exponential distribution)~\cite{BV2014}: 
\begin{align}
\mathcal{H}(P) & = P \exp\left[-s\mathcal{Y}^{1/2}\right], \\
f(r) & = 1-\dfrac{2M}{r}\exp\left(-\frac{Q^2}{2 Mr}\right),\\
\phi(r) & = \dfrac{3M}{2Q}+\dfrac{1}{4Qr}\exp\left(-\frac{Q^2}{2 Mr}\right)\left(Q^{2}-6Mr\right).
\end{align}
\end{itemize}
For simplicity, we defined $\mathcal{Y} = \sqrt{-Q^{2}P/2}$, $s = |Q|/2M$, $\alpha = \pi^{2}Q^{3}/64\sqrt{2}M^{2}$, and $z \equiv \pi Q^{2}/8M$. The parameters $Q$ and $M$ are associated with the charge and mass of the objects, respectively. Moreover, we also remark that the BV model obtained by means of an exponential distribution in NED was also derived as an anisotropic fluid, see Refs.~\cite{HC2013,HC2015}.

\section{SUFFICIENT CONDITIONS FOR ELECTRICALLY CHARGED BHs DERIVED IN THE $F$ FRAMEWORK}\label{Appendix}

We considered BHs derived in the $P$ framework, as in this context we can obtain electrically charged RBHs in a simpler way. Nevertheless, it is also possible to get electrically charged BHs in the $F$ framework, although they cannot be simultaneously regular and have a Maxwell-like behavior at infinity for just one Lagrangian density $\mathcal{L}(F)$.

In the $F$ framework, the electric field is given by
\begin{equation}
\label{ef2}E(r) = \dfrac{Q}{r^{2}\mathcal{L}_{F}},
\end{equation}
and then the corresponding radial electrostatic potential yields
\begin{equation}
\label{pott}\phi(r) = -\int_{\infty}^{r}\dfrac{Q}{x^{2}\mathcal{L}_{F}}dx.
\end{equation}

The analogue of Eq.~\eqref{ratio2} can be written as
\begin{align}
\nonumber \Phi_{\rm{hor}} = & \ \dfrac{Q^{\rm{mod}}}{Q^{\rm{RN}}}\dfrac{r_{+}^{\rm{RN}}}{r_{+}^{\rm{mod}}}\left(\dfrac{d \mathcal{L}}{dF} \right)^{-1}\bigg|_{r = r_{+}} +\dfrac{r_{+}^{\rm{RN}}}{Q^{\rm{RN}}}\times \\
\label{ratio3}&\left[\int_{0}^{\mathcal{F}_{+}}\left(\dfrac{-Q^{2}\mathcal{L}_{F}^{2}F}{2}\right)^{1/4}\left(\dfrac{\mathcal{L}_{FF}}{\mathcal{L}_{F}^{3}\Phi}\right)d\left( \mathcal{L}_{F}^{2}F\right)\right]_{\rm{mod}},
\end{align}
where $\mathcal{F}_{+} = \left(-2Q^2/\mathcal{L}_{F}^{2}F\right)^{1/4}\big|_{+}$ and $\Phi \equiv \mathcal{L}_{F} +2F\mathcal{L}_{FF}$. As we can see, this expression is very complicated and it is difficult to impose conditions over the quantities associated with the NED, to guarantee $\Phi_{\rm{hor}} > 1$.

\end{appendix}

\end{document}